\documentclass[journal]{IEEEtran}

\usepackage{cite}
\usepackage[cmex10]{amsmath}
\usepackage{amsfonts}
\usepackage{algorithm}
\usepackage{epsfig}
\usepackage[noend]{algpseudocode}
\usepackage{bm}
\usepackage{booktabs}
\usepackage{color}
\usepackage[T1]{fontenc}
\usepackage{url}
\usepackage{leftindex}
\usepackage{listings}
\usepackage{hyperref}

\usepackage{xr}
\makeatletter
\newcommand*{\addFileDependency}[1]{
	\typeout{(#1)}
	\@addtofilelist{#1}
	\IfFileExists{#1}{}{\typeout{No file #1.}}
}

\addFileDependency{supplement.aux}%

\newcommand{\isdef}{\overset{\text{def}}{=}}
\newcommand{\norm}[1]{\left\lVert {#1} \right\rVert}
\newcommand{\abs}[1]{\lvert {#1} \rvert}
\newcommand{\compslc}[3]{\kappa^{(#1)}_{{#2}\to{#3}} }

\begin{document}
\bstctlcite{IEEEexample:BSTcontrol}

\title{Near-Real-Time InSAR Phase Estimation for Large-Scale Surface Displacement Monitoring}

\author{Scott~Staniewicz,
Sara~Mirzaee,
Heresh~Fattahi,
Talib~Oliver-Cabrera,
Emre~Havazli,
Geoffrey~Gunter,
Se-Yeon~Jeon,
Mary~Grace~Bato,
Jinwoo~Kim,
Simran~S.~Sangha,
Bruce~Chapman,
Alexander~L.~Handwerger,
Marin~Govorcin,
Piyush~Agram,
David~Bekaert
\thanks{S.~Staniewicz, S.~Mirzaee, H.~Fattahi, T.~Oliver-Cabrera, E.~Havazli, G.~Gunter, S.~Jeon, M.G.~Bato, S.~Sangha, A.~Handwerger, M.~Govorcin, D.~Bekaert are with the Jet Propulsion Laboratory, California Institute of Technology.}%
\thanks{P. Agram is with EarthDaily Analytics.}%
\thanks{J. Kim is with Southern Methodist University.}%
\thanks{D. Bekaert is with VITO, Flemish Institute for Technological Research, Mol, Belgium}%
\thanks{
  © 2025. All rights reserved.}
}

\maketitle

\begin{abstract}
    Operational near-real-time monitoring of Earth's surface deformation using Interferometric Synthetic Aperture Radar (InSAR) requires processing algorithms that efficiently incorporate new acquisitions without reprocessing historical archives. We present a sequential phase linking approach using compressed single-look-complex images (SLCs) capable of producing surface displacement estimates within hours of the time of a new acquisition. Our key algorithmic contribution is a mini-stack reference scheme that maintains phase consistency across processing batches without adjusting or re-estimating previous time steps, enabling straightforward operational deployment. We introduce online methods for persistent and distributed scatterer identification that adapt to temporal changes in surface properties through incremental amplitude statistics updates. The processing chain incorporates multiple complementary metrics for pixel quality that are reliable for small SLC stack sizes, as well as an $L_1$-norm network inversion to limit propagation of unwrapping errors across the time series. We use our algorithm to produce the OPERA Surface Displacement from Sentinel-1 (DISP-S1) product, the first continental-scale surface displacement product over North America. Validation against GPS measurements and InSAR residual analysis demonstrates millimeter-level agreement in velocity estimates in varying environmental conditions. We also demonstrate our algorithm's capabilities with a successful recovery of meter-scale co-eruptive displacement at Kilauea volcano during the 2018 eruption, as well as detection of subtle uplift at Three Sisters volcano, Oregon--- a challenging environment for C-band InSAR due to dense vegetation and seasonal snow. We have made all software available as open source libraries, providing a significant advancement to the open scientific community's ability to process large InSAR data sets in a cloud environment.
\end{abstract}


\IEEEpeerreviewmaketitle

\section{Introduction}\label{sec:introduction}
\IEEEPARstart{T}{he} operational monitoring of changes to Earth's surface has emerged as a critical capability for hazard assessment, infrastructure management, and many other geophysical applications \cite{Rosen2000SyntheticApertureRadar,Burgmann2000SyntheticApertureRadar,Crosetto2016PersistentScattererInterferometry,Raspini2018ContinuousSemiautomaticMonitoring}. While the Sentinel-1 mission \cite{Torres2012GMESSentinel1Mission} has provided unprecedented data availability to map surface displacements using the interferometric synthetic aperture radar (InSAR) technique \cite{Massonnet1998RadarInterferometryIts,Bamler1998SyntheticApertureRadar}, translating this wealth of data into timely, user-friendly products remains challenging. The NASA Observational Products for End-Users from Remote Sensing Analysis (OPERA) project \cite{opera_project,fattahi2022opera} aims to address this gap by producing the Surface Displacement from Sentinel-1 (DISP-S1) product at 30-meter posting and low latency for a North America\footnote{North America corresponds to the United States (USA) and U.S. Territories, parts of Canada within 200 km of the U.S. border, and all mainland countries from the southern US border up to and including Panama} geographic scope. These demanding specifications require improvements to the underlying estimation algorithms to produce low-latency products with high reliability. In this paper, we develop the algorithmic approach for the DISP-S1 products, as well as the foundation for the upcoming NISAR-based surface displacement products (DISP-NISAR).

The feasibility of large-scale InSAR monitoring has been demonstrated by several recent regional \cite{ Cigna2022LandSubsidenceAquiferSystem,Thollard2021FLATSIMForMTerLArgeScale,Lazecky2020LiCSARAutomaticInSAR} and nation-wide \cite{Dehls2019INSARNoNational,Costantini2021EuropeanGroundMotion} ground displacement services. However, these products update on a fixed basis (yearly in the case of the European Ground Motion Service (EGMS) \cite{Crosetto2023PanEuropeanDeformation}) limiting their applications. In order to provide near-real-time (NRT) latency displacement products, where updates are created shortly after one or several new single-look complex (SLC) images are acquired without reprocessing the full historical archive of data, new algorithms are required that are able to run on a continental scale.

Phase linking, an advanced InSAR time series technique for estimating wrapped interferometric phase \cite{Guarnieri2008ExploitationTargetStatistics,Tebaldini2010MethodsPerformancesMultiPass,Ferretti2011NewAlgorithmProcessing}, offers a promising foundation for NRT processing which can avoid the cumulative errors arising from sequential unwrapping \cite{Chen2001TwodimensionalPhaseUnwrapping,Mirzaee2023NonlinearPhaseLinking}. Unlike the Small Baseline Subset (SBAS) approach \cite{Berardino2002NewAlgorithmSurface}, which requires careful interferogram network selection to maintain coherence \cite{Pepe2005LimitationsSmallBaseline}, phase linking uses the full covariance matrix from a stack of coregistered single-look complex (SLC) images to estimate an optimized wrapped time series for Distributed Scatterers (DS) \cite{Ferretti2011NewAlgorithmProcessing,Guarnieri2008ExploitationTargetStatistics}.
Our approach builds directly on the sequential phase linking framework of Ansari et al. \cite{Ansari2017SequentialEstimatorEfficient}, which introduced efficient batch processing using mini-stacks'' of SLCs and  ``compressed SLCs'' which summarize historical information \cite{Ansari2017SequentialEstimatorEfficient}. We adopt the core formulation for phase estimation from \cite{Mirzaee2023NonlinearPhaseLinking} and covariance compression \cite{Ansari2017SequentialEstimatorEfficient}, leveraging their theoretical analysis of estimator performance. However, their formulation requires an adjustment step (i.e., ``datum adjustment'' ), which introduces extra complexity and may be difficult for products archived at a Distributed Active Archive Center (DAAC). Hence, notable challenges still exist for creating a wide-area, operationally-friendly time series algorithm and implementation.

In this work, we present a comprehensive NRT processing chain for the OPERA DISP-S1 product that addresses these challenges via algorithm and implementation optimizations. The OPERA project, managed by the Jet Propulsion Laboratory, California Institute of Technology, produces analysis ready data products that meet the end-user needs of U.S. Federal Agencies (\url{https://www.jpl.nasa.gov/go/opera/}).
Our key contributions include (1) a novel sequential phase linking formulation which maintains consistency across mini-stacks without datum adjustments; (2) online methods for persistent scatter (PS) \cite{Ferretti2001PermanentScattersSAR,Hooper2004NewMethodMeasuring} and DS identification that adapt to temporal changes in surface scattering properties; (3) quality metrics designed for small mini-stack sizes (under 30 images) that provide reliable pixel selection; (4) a sensor-agnostic, open-source implementation \cite{Staniewicz2024DolphinPythonPackage} with fully-reproducible data products.
In Section \ref{sec:methods}, we outline our sequential phase estimation approach. Section \ref{sec:proc-details} details the complete processing chain, including unwrapping and time series inversion. Section \ref{sec:processing-config} describes specific the input data and configuration used to create the DISP-S1 products, and Section \ref{sec:validation-methods} outlines the OPERA project validation approach.
Finally, validated line-of-sight displacement results across a diverse set of environments are shown in Section \ref{sec:results}, demonstrating the system's operational readiness.

\section{Methods}
\label{sec:methods}

\subsection{Phase Linking Background}
\label{subsec:methods-phase-linking-background}

Phase linking estimates the wrapped phase evolution of a pixel from a stack of $N$ coregistered SLC SAR images \cite{Guarnieri2008ExploitationTargetStatistics}.
Let $\mathbf{z} \in \mathbb{C}^{N}$ be a vector of complex pixel values for a particular spatial location.
We assume $\mathbf{z}$ arises from many independent scatterers in the resolution element (i.e., a DS) and follows a circular complex Gaussian (CCG) distribution \cite{Bamler1998SyntheticApertureRadar}:
\begin{equation}
  \mathbf{z} \sim \mathcal{CN}(\mathbf{0}, \Sigma)  = \frac{1}{\pi^N \det (\Sigma)} \exp \left[-\mathbf{z}^H \Sigma^{-1} \mathbf{z} \right],
\end{equation}
where $\Sigma = \mathbb{E}[\mathbf{z} \mathbf{z}^{H}] $ is the covariance matrix, $(\cdot)^{H}$ denotes the Hermitian (conjugate transpose) operator, and $\det$ is the matrix determinant.
For convenience, we assume each component of $\mathbf{z}$ has unit average power ($\mathbb{E}[|\mathbf{z}_i|^2] = 1$) so that $\Sigma $ can be treated as a coherence matrix whose magnitudes range from 0 to 1 \cite{Cao2015MathematicalFrameworkPhaseTriangulation}.
We can decompose $\Sigma$ into a product of the unknown ``true'' phases and coherence magnitudes:
\begin{equation}
  \Sigma = \Phi^{H} \Gamma \Phi,
\end{equation}
where $\Phi = \mathrm{diag} \left(e^{j \boldsymbol{\phi}}\right)$ is a diagonal matrix of unit-magnitude complex values, $\boldsymbol{\phi} = [\phi_{1}, \dots, \phi_{N}]$ is vector of ``true'' phases to be estimated, and $\Gamma$ is the real-symmetric ${N \times N}$ matrix of normalized coherence magnitudes $\gamma_{mn}$, representing correlations among the $m$th and $n$th SAR images. In practice, we estimate $\Sigma$ by averaging $L$ samples in a statistically homogeneous pixel (SHP) neighborhood $\Omega$:
\begin{equation}
  \hat{\Sigma} = \frac{1}{L} \sum_{l=1}^L \mathbf{z}^{(l)} \bigl(\mathbf{z}^{(l)}\bigr)^{H}. \label{eq:sample-coherence}
\end{equation}

The goal of phase linking is to recover a consistent set of optimal phase values $\boldsymbol{\phi}$ from $\hat{\Sigma}$. The log-likelihood of the data $\log L(\Sigma)$ is proportional to
\begin{equation*}
  \log  L(\Sigma) \isdef \log p(\mathbf{z} \mid \Sigma) \propto \text{trace}\left(-\Phi^{H} \Gamma^{-1} \Phi \hat{\Sigma}\right).
\end{equation*}
Let $\boldsymbol{\zeta} = [e^{j\phi_1}, \dots, e^{j\phi_N}]$ denote the vector of unit-magnitude complex values corresponding to the phases such that $\Phi = \text{diag}(\boldsymbol{\zeta})$.
Using the matrix identity
$\operatorname{trace}\!\left(\Phi^{H}A\Phi B\right) 
= \boldsymbol{\zeta}^{H}\!\big(A\!\circ\! B^{H}\big)\boldsymbol{\zeta}$, 
where $\circ$ denotes element-wise multiplication (Hadamard product), 
we can rewrite $\Phi^{H}\!\Gamma^{-1}\!\Phi\hat{\Sigma}$ as 
$\boldsymbol{\zeta}^{H}\!\big(\Gamma^{-1}\!\circ\!\hat{\Sigma}\big)\boldsymbol{\zeta}$ \cite{Horn2012MatrixAnalysisSec75, Guarnieri2008ExploitationTargetStatistics}.
When $\Gamma$ is treated as known, the maximum likelihood estimate of the phase, $\hat{\boldsymbol{\phi}}$, solves the optimization problem \cite{Guarnieri2008ExploitationTargetStatistics,Ferretti2011NewAlgorithmProcessing}
\begin{equation}
\begin{aligned}
	& \underset{\boldsymbol{\zeta}}{\text{minimize}} & & 	\boldsymbol{\zeta}^{H}\!\Big(\Gamma^{-1}\!\circ\!\hat{\Sigma}\Big)\boldsymbol{\zeta} \\ 
	& \text{subject to} & & |\zeta_i|=1~\forall i. \label{eq:phase-linking-optimization}
\end{aligned}
\end{equation}
We then extract the phase estimates as $\hat{\boldsymbol{\phi}} = \angle  \boldsymbol{\zeta}$.
Since the true coherence matrix $\Gamma$ is unknown, the sample coherence $\abs{\hat{\Sigma}}$ is often used in place of $\Gamma$, where $\abs{\cdot}$ denotes element-wise absolute value.

Because interferometry measures only phase differences, one of the $N$ phases in $\boldsymbol{\phi}$ is not independently determinable \cite{Tebaldini2010MethodsPerformancesMultiPass}. A reference index must be chosen:
\begin{equation*}
  \zeta_r \isdef  e^{j\phi_r}.
\end{equation*}
The reference operation is performed by multiplying all elements of $ \hat{\boldsymbol{\zeta}}$ by $\zeta_r^{*}$, which sets $\hat{\phi}_r = 0$. A common choice is the first date, $r = 1$ \cite{Ansari2018EfficientPhaseEstimation,Tebaldini2010MethodsPerformancesMultiPass}, but other reference schemes (e.g., last date) can offer advantages in sequential processing (See Supplement \ref{sec:supplement-sequential-last-per-ministack}).

Note that by choosing a reference phase, we are effectively estimating single-reference interferograms. The phase from the reference index is contained in all phase linking outputs; however, shorter temporal baseline interferograms may later be re-formed by simple cross-multiplication, since the reference phase cancels out.

\subsection{Batched processing and compressed SLCs}\label{subsec:methods-sequential}

\begin{figure*}[!ht]
	\centering
	\includegraphics[width=0.95\textwidth]{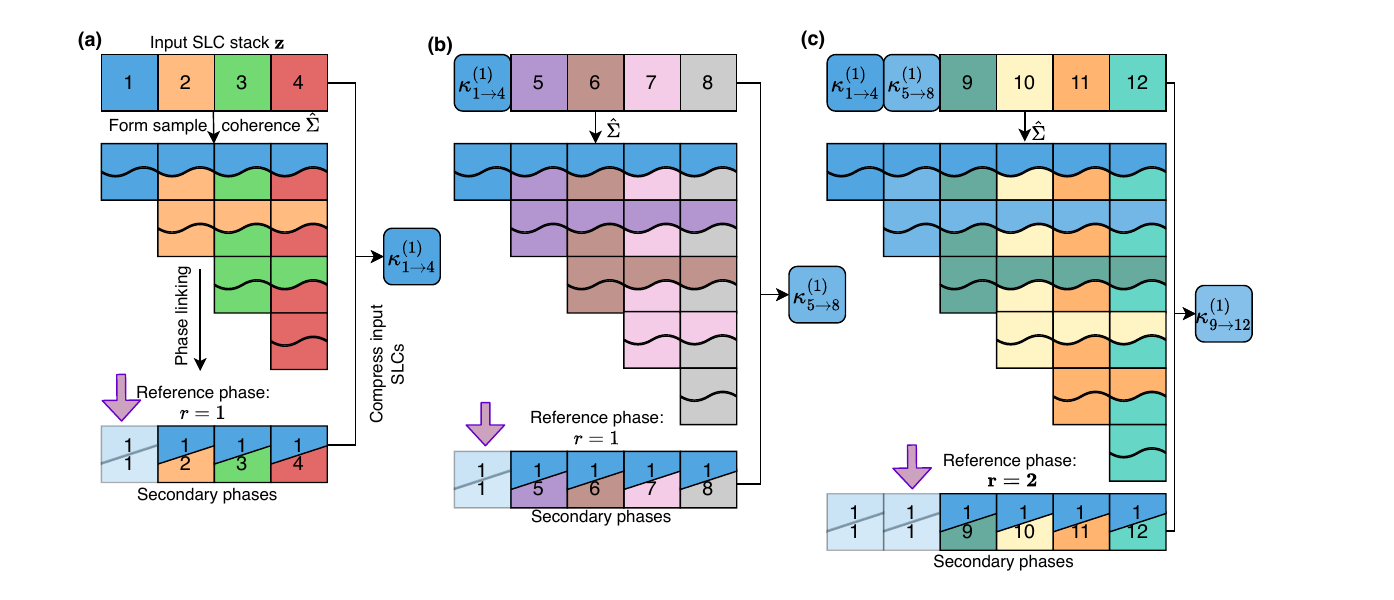}
	\caption{Batch processing of $N=12$ SLC images in mini-stacks of size $M=4$. Squares indicate input SLCs, wavy lines indicate multi-looked interferograms, rounded squares indicate compressed SLCs.
		(a) The first mini-stack (SLC images 1-4) is phase-linked with reference index $r=1$. Input SLCs are compressed to produce $\compslc{1}{1}{4}$, and interferograms are formed with SLC 1 as reference.
		(b) The second mini-stack uses $\compslc{1}{1}{4}$ as a reference to create $\compslc{1}{5}{8}$ and to produce interferograms relative to day 1.
		(c) Subsequent mini-stacks use the most recently formed compressed SLC as reference. For mini-stack 3, this means that the mini-stack uses reference $r=2$, which is $\compslc{1}{5}{8}$. The outputs are interferograms relative to day 1. }
	\label{fig:sequential-demo-always-first}
\end{figure*}

Modern SAR missions (e.g., Sentinel-1) may acquire hundreds of images over a given region, which makes processing the full $N \times N$ covariance matrix computationally expensive. One strategy is to partition the time series into smaller batches of size $M$, called \emph{mini-stacks} \cite{Ansari2017SequentialEstimatorEfficient}.
Each mini-stack is processed to obtain a phase-linked estimate, which is then used to form a \emph{compressed} SLC, distilling information in the mini-stack into a single complex image \cite{Zan2007PsProcessingDecorrelating}.

Let $\hat{\boldsymbol{\zeta}} = e^{j \hat{\boldsymbol{\phi}} } $ be the phase-linking solution for a mini-stack of $M$ images. At each pixel, the compressed SLC pixel $\kappa \in \mathbb{C}$ is formed by projecting the original SLC vector $\mathbf{z}$ onto the phase-linking solution:
\begin{equation*}
  \kappa = \langle \hat{\boldsymbol{\zeta}}, \mathbf{z} \rangle = \hat{\boldsymbol{\zeta}}^{H} \mathbf{z}.
\end{equation*}
Writing $z_i = a_i \, e^{j\phi_i}$ and $\hat{\zeta}_i = e^{j \hat{\phi}_i}$, we see that any chosen reference phase factors out of this inner product:
\begin{align*}
  \kappa
  &= \sum_{i=1}^M a_i \, e^{j\phi_i} \, \bigl(e^{j \hat{\phi}_i}\bigr)^{*}\,e^{j \hat{\phi}_r} \\
  &= e^{j \hat{\phi}_r} \sum_{i=1}^M a_i \, e^{j(\phi_i - \hat{\phi}_i)}.
\end{align*}
Thus, each compressed SLC pixel has a ``base phase'' from the reference $\hat{\phi}_r$, plus a weighted sum of the residuals from the phase linking solution.
We can label a compressed SLC with reference phase $\hat{\phi}_r$ created from the mini-stack containing images $i,i+1,\dots, i+M$ as $ \compslc{r}{i}{i+M} $.

In our sequential batch processing scheme, the compressed SLC from one mini-stack is prepended to the next mini-stack, increasing each mini-stack's size by one. Compressed SLCs allow the formation of longer interferograms (reducing displacement bias from short-lived phase signals like soil moisture \cite{Ansari2021StudySystematicBias}) but reduce the total computational cost by shrinking the size of the covariance matrix.

\paragraph*{Proposed Approach}
The processing scheme outlined in Figure~\ref{fig:sequential-demo-always-first} produces a set of single-reference interferograms relative to the first input date by referencing mini-stacks to a previously generated compressed SLC:

\begin{enumerate}
  \item \emph{First Mini-stack} ($\{1,\dots,M\}$) (Figure~\ref{fig:sequential-demo-always-first}a):
    \begin{enumerate}
      \item Perform phase linking to get $\hat{\boldsymbol{\zeta}}^{(1)}$.
      \item Form compressed SLC $\compslc{1}{1}{M}$.
    \end{enumerate}
  \item \emph{Second Mini-stack} ($\{\compslc{1}{1}{M}, z_{M+1}, \dots, z_{2M}\}$) (Figure~\ref{fig:sequential-demo-always-first}b):
    \begin{enumerate}
      \item Use $\compslc{1}{1}{M}$ as the reference image (since it is already phase-referenced to the first mini-stack).
      \item After phase linking, form the new compressed SLC $\compslc{1}{M+1}{2M}$ by projecting $\{ z_{M+1},\dots,z_{2M}\}$ onto the new solution $\hat{\boldsymbol{\zeta}}^{(2)}$.
    \end{enumerate}
  \item \emph{Third Mini-stack} ($\{\compslc{1}{1}{M}, \compslc{1}{M+1}{2M}, z_{2M+1}, \dots, z_{3M}\}$) (Figure~\ref{fig:sequential-demo-always-first}c):
    \begin{enumerate}
      \item Use $\compslc{1}{M+1}{2M}$ as the reference image
      \item After phase linking, form the new compressed SLC $\compslc{1}{2M+1}{3M}$
    \end{enumerate}
\end{enumerate}

We note that $\compslc{1}{M+1}{2M}$ ``inherits'' the reference phase from its reference image $\compslc{1}{1}{M}$. This allows the second mini-stack, and all subsequent mini-stacks, to directly exploit the ``base phase'' from day~1 and skip any re-referencing (datum adjustment) steps (c.f. \cite{Ansari2017SequentialEstimatorEfficient}). This also lets us use only the $K$ most recent compressed SLCs, even for large (>300) stacks of SLCs, limiting the maximum computational time needed for phase linking.

\subsection{Demonstration with synthetic data}
\label{subsec:synthetic-ifg-demo}

\begin{figure}[!ht]
	\centering
	\includegraphics[width=.95\columnwidth]{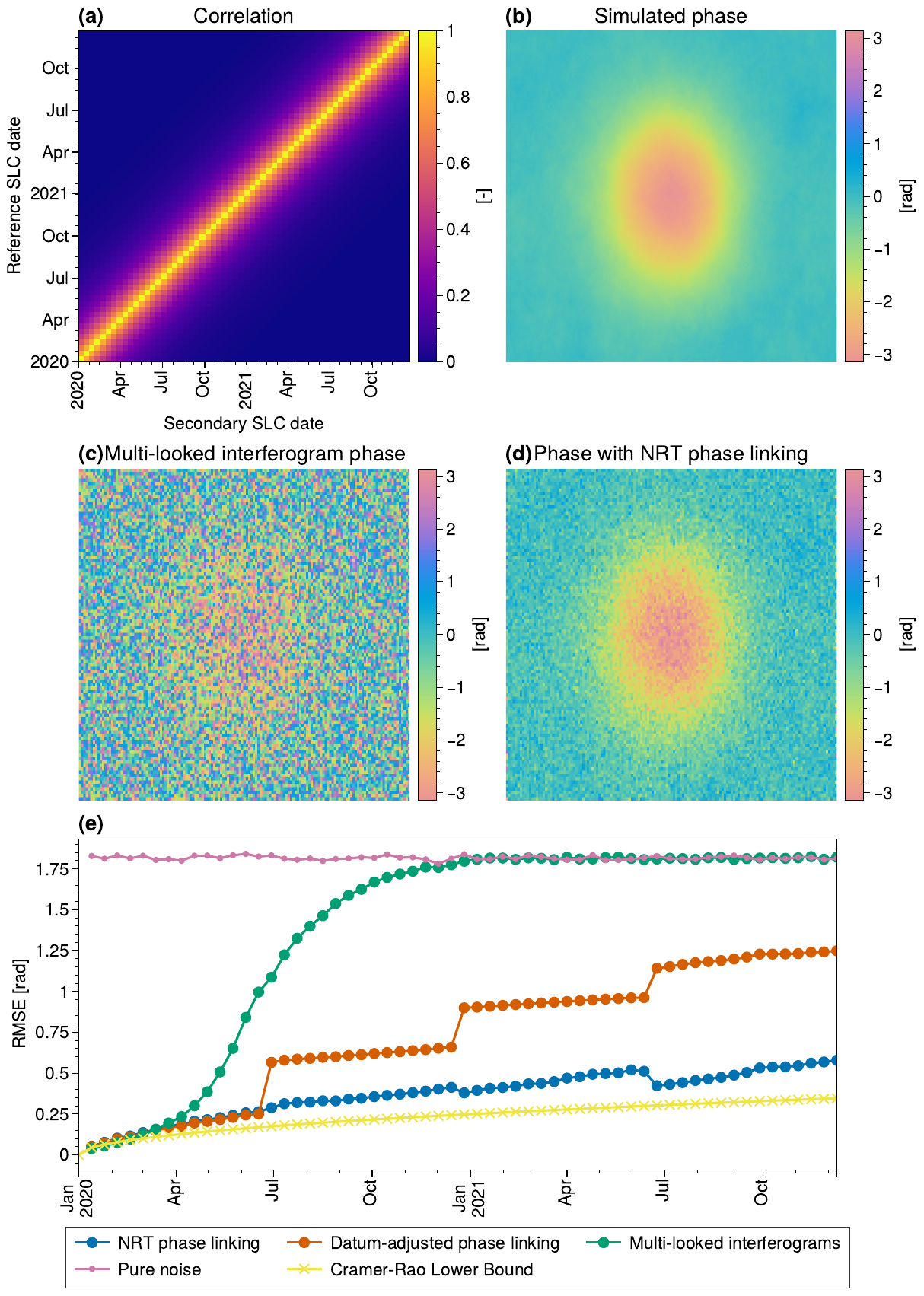}
	\caption{Synthetic demonstration of sequential phase linking for a $1500\times1500$ pixel scene, processed in mini-stacks of 15~SLCs. 
		(a) Simulated temporal correlation matrix. 
		(b) Example of simulated ``true'' phase, consisting of a deformation bowl plus atmospheric noise.
		(c) Conventional interferogram formed from two real SLCs ($s_1 s_{20}^*$) after multi-looking $(15,15)$.
		(d) Interferogram for the same time span using the compressed SLC $\compslc{1}{1}{15}$ as the reference, showing improved coherence. 
		(e) Comparison of RMSE versus time for different estimators: near-real-time phase linking (blue), datum-adjusted phase linking (orange), multi-looked interferograms (green), pure noise (pink), and the Cram\'er–Rao lower bound (yellow).}
	\label{fig:synthetic-example}
\end{figure}

To illustrate the sequential phase linking algorithm and the effect of compressed SLC formation, we generated a stack of synthetic SLCs with exponentially decaying temporal correlation (Figure~\ref{fig:synthetic-example}). The noise in each SLC was drawn from a circular complex Gaussian distribution with correlation 
\[
\rho(t) = (\rho_0 - \rho_{\infty}) e^{-t / \tau} + \rho_{\infty},
\]
where $\rho(t)$ is the correlation at temporal baseline $t$, $\rho_0$ is the initial correlation, $\tau$ is the exponential decay constant, and $\rho_{\infty}$ is the long-term coherence. We used $\rho_0 = 1$, $\rho_{\infty} = 0$, $\tau = 60$ days, and simulated 60 SLCs sampled every 12 days (Figure~\ref{fig:synthetic-example}a). The simulated ``true'' phase of each SLC was modeled as minor turbulent tropospheric noise \cite{Hanssen2001RadarInterferometryData} and a deformation uplifting bowl with a linear rate of 5 radians per year (Figure~\ref{fig:synthetic-example}b).

We processed the data using mini-stacks of 15 images, following the algorithm in Section~\ref{subsec:methods-sequential}. The improvement gained from using the compressed SLCs is visible by comparing a conventional multi-looked interferogram (Figure~\ref{fig:synthetic-example}c) to one formed using the compressed SLC as the reference (Figure~\ref{fig:synthetic-example}d). The latter exhibits much higher coherence while preserving the same mean phase, showing that the mini-stack filtering effectively suppresses decorrelation noise.

Figure~\ref{fig:synthetic-example}e quantifies the performance of different estimators. The NRT sequential phase linking (blue) closely approaches the Cramer-Rao lower bound (CRLB, yellow, \cite{Guarnieri2007HybridCramErRao}), while multi-looked interferograms (green) and pure noise (pink) deviate substantially after about 30~days of decorrelation. The ``datum-adjusted'' sequential phase linking (orange), following the approach of \cite{Ansari2017SequentialEstimatorEfficient}, yields comparable results but requires an additional optimization step to align phase offsets between consecutive mini-stacks. Our NRT approach integrates this referencing during each mini-stack, eliminating the need for post-processing adjustments.

\section{Near-real-time InSAR time-series estimation}
\label{sec:proc-details}

\begin{figure*}[!ht]
	\centering
	
	\includegraphics[width=0.9\textwidth]{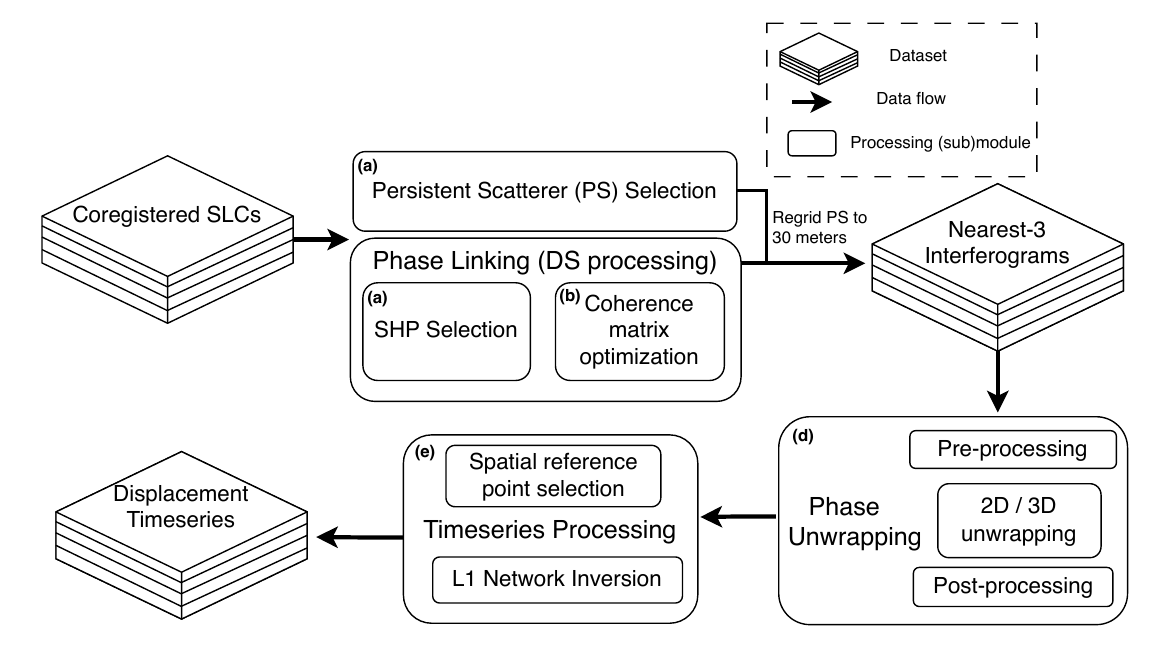}
	\caption{Summary of the modules used in the displacement processing workflow. Letters refer to subsections of Section \ref{sec:proc-details} providing details.}
	\label{fig:processing-modules}
\end{figure*}

Here we describe details of our wide-area, near-real-time InSAR processing chain (Figure \ref{fig:processing-modules}).
The NRT context imposes two key constraints on our approach:
\begin{enumerate}
  \item We must be able to incorporate new acquisitions without accessing the full historical archive.
  \item Processing speed must be considered due to computational costs, which favors processing smaller subsets of SLCs when possible.
\end{enumerate}
These constraints particularly affect how we estimate metrics for PS and DS selection, as well as how we maintain reliable quality statistics using limited numbers of images.

\subsection{Online Persistent and Distributed Scatterer Selection}
\label{subsec:ps-shp-selection}

Our algorithm identifies both PS and SHP pixel neighborhoods for DS using amplitude-based tests. PS candidates are selected using the amplitude dispersion index
\begin{equation}
  D_A \isdef \frac{\sigma_A}{\mu_A},
\end{equation}
where $\mu_A$ and $\sigma_A$ are the mean and standard deviation of the pixel amplitude across available acquisitions.

To process DS, we first identify an adaptive SHP neighborhood $\Omega$ around each pixel with the Generalized Likelihood Ratio Test (GLRT) \cite{Parizzi2011AdaptiveInSARStack}.  The GLRT requires only the amplitude means and variances, making it computationally comparable to simple rectangular-window averaging. However, in the NRT context, two challenges arise:

\begin{enumerate}
  \item The mean and standard deviation of SAR backscatter amplitude estimated from small sample sizes may be noisy.
  \item Temporal variation of surface scattering properties may lead to changes in the amplitude of the SAR backscatter.
\end{enumerate}

We address these challenges by storing mini-stack amplitude means and variances as ancillary data products along with each newly formed compressed SLC. We then merge the old and new amplitude statistics through standard formulas for combining group means and variances:
\begin{align}
  \mu_{\text{new}} &= \frac{\sum_i w_i\,\mu_i}{\sum_i w_i}, \label{eq:merged-mean} \\
  \sigma_{\text{new}}^2 &= \frac{\sum_i w_i(\sigma_i^2 + \mu_i^2)}{\sum_i w_i} \;-\; \bigl(\mu_{\text{new}}\bigr)^2, \label{eq:merged-variance}
\end{align}
where $\mu_i$, $\sigma_i^2$, and $w_i$ are the amplitude mean, amplitude variance, and weight of each group. Two simple options for $w_i$ are to evenly weight the previous $n$ images, or to use an exponential decay to prioritize recent observations, while still allowing statistics to adapt to changes.

\subsection{Phase Linking Method}
\label{subsec:proc-phase-linking}

Once an adaptive SHP neighborhood $\Omega$ is identified for each pixel, we form the sample coherence matrix \eqref{eq:sample-coherence} averaging over pixels in $\Omega$ (Figure \ref{fig:processing-modules}b).
Our phase linking procedure follows the Combined Phase Linking approach from \cite{Mirzaee2023NonlinearPhaseLinking}.
We solve the maximum likelihood phase estimation problem using EMI (Eigendecomposition-based Maximum-likelihood estimator of Interferometric phase \cite{Ansari2018EfficientPhaseEstimation}).
EMI efficiently solves Equation \eqref{eq:phase-linking-optimization} as an eigenvalue problem
\begin{equation}
  \left( \hat{\Gamma}^{-1} \circ \hat{\Sigma}\right) \hat{\boldsymbol{\zeta}} = \lambda_{\min} \hat{\boldsymbol{\zeta} }  \label{eq:emi-eigenvalue-problem},
\end{equation}
where $|\hat{\Sigma}|^{-1}$ is plugged in for $\hat{\Gamma}^{-1}$. The phase of the eigenvector corresponding to the smallest eigenvalue of $( \hat{\Gamma}^{-1} \circ \hat{\Sigma} )$ gives the phase-linking solution.
At pixels where the inversion of $\hat{\Gamma}$ fails due to numerical issues, we revert to the Eigenvalue decomposition (EVD) method: we use the phase of the eigenvector corresponding to the largest eigenvalue of the sample coherence matrix $\hat{\Sigma}$ \cite{Fornaro2015CAESARApproachBased}.
Note that the inversion is more likely to fail at very high coherence pixels where many coherence magnitudes are near 1; thus, the fallback to EVD at these pixels still provides a high-quality phase solution.

Since we only need a single eigenvector (corresponding to $\lambda_{\min}$), the inverse power method \cite{Ipsen1997ComputingEigenvectorInverse} is fast and effective for our typical small problem sizes of $M \approx 15-30$.
Noting that eigenvalues of $\Gamma^{-1} \circ \Sigma$ all have magnitude greater than or equal to $1$ (\cite{Ansari2018EfficientPhaseEstimation} Appendix B), we set initial guess $\lambda_{\text{min}}^{(0)} = 0.99$. Since the dominant eigenvalue is close to 1, with values slightly above 1 in lower coherence regions, this choice ensures rapid convergence (typically under 10 iterations).

\subsection{Quality metrics: Temporal coherence and phase similarity}
\label{subsec:methods-similarity}

The quality of phase-linked DS pixels is commonly assessed using the temporal coherence $\gamma_{t}$ \cite{Cao2015MathematicalFrameworkPhaseTriangulation}, defined as:
\begin{equation}
  \label{eq:temporal-coherence}
  \gamma_{t}
  \;=\;
  \frac{1}{N_{\text{ifg}}}
  \left\lvert \sum_{i=1}^M
  \sum_{k=i+1}^M
  \exp\Bigl\{\, j \bigl(\phi_{i,k} - \hat{\phi}_{i,k}\bigr)\Bigr\}
  \right\rvert,
\end{equation}
where $N_{\text{ifg}} = \frac{M(M-1)}{2}$ is the total number of interferograms, and $\phi_{i,k}$ and $\hat{\phi}_{i,k}$ represent the phase differences between the $i$th and $k$th images for the original multi-looked interferogram phase and phase linking solution, respectively.
This metric, also known as ``goodness of fit'' $\gamma_{\text{PTA}}$ \cite{Ferretti2011NewAlgorithmProcessing} or ``a posteriori coherence'' $\gamma_{\text{apt}}$ \cite{Ansari2018EfficientPhaseEstimation}, quantifies how well the estimated phases match the original observations.

While $\gamma_t$ is effective for batch processing, it has limitations when applied to smaller mini-stacks.
For $M=2$ images, it always yields $\gamma_t = 1$, and for $M < 20$, its limited dynamic range makes it difficult to distinguish between high- and low-quality pixels. For more on the effect of mini-stack size on $\gamma_t$, see Section \ref{sec:discussion-ministack-size}.

To address these limitations, we introduce a complementary metric based on the cosine similarity measure \cite{Wang2022AccuratePersistentScatterer}. For two vectors $\mathbf{x}$ and $\mathbf{y}$ of length $K$, the real-valued cosine similarity is:
\begin{equation}
  s_{\mathbf{x}, \mathbf{y}}
  \;=\;
  \frac{1}{K} \sum_{i=1}^K
  \cos\bigl(x_i - y_i\bigr).
\end{equation}

We extend this concept to InSAR phase data by considering a search window $W$ of radius $r$ (typically $\sim 200$\,m) around each pixel.
For a given pixel with phase-linked estimate $\hat{\boldsymbol{\phi}}$, we compute its phase similarity with each neighboring pixel $\mathbf{x}$ in $W$:
\begin{equation}
  s_{\hat{\boldsymbol{\phi}},\,\mathbf{x}}
  \;=\;
  \frac{1}{N_{\text{ifg}}}
  \sum_{i=1}^{N_{\text{ifg}}}
  \cos\Bigl(\,\hat{\phi}_i \;-\; x_i\Bigr),
\end{equation}
where $x_i$ is the phase of image $i$ at the given neighboring pixel, and ${N_{\text{ifg}}}$ is the number of re-formed interferograms that will be unwrapped (typically a nearest-3 network, see Section \ref{subsec:methods-phase-unwrapping}). The phase cosine similarity $\gamma_{s}$ is then defined as the median of these similarities:
\begin{equation}
  \label{eq:phase-cosine-sim}
  \gamma_{s}
  \;=\;
  \mathrm{median}_{\mathbf{x}\in W}
  \Bigl\{
    s_{\hat{\boldsymbol{\phi}},\,\mathbf{x}}
  \Bigr\}.
\end{equation}

The $\gamma_s$ metric measures how well a pixel's optimized phase history matches the neighboring pixels in space. Higher values indicate better agreement with local phase patterns. The $\mathrm{median}$ operation makes it more robust even with small SLC stack sizes, and it provides a measure of pixel quality complementary to $\gamma_t$. 

\subsection{Phase Unwrapping}
\label{subsec:methods-phase-unwrapping}

After obtaining wrapped phases from phase linking, we unwrap the phase to recover the continuous deformation signal (Figure \ref{fig:processing-modules}d). By default, we create a network of $N_{\text{ifg}} = 3M - 6$ interferograms using a nearest-3 neighbor network, under the assumption that the perpendicular baselines of Sentinel-1 are all relatively small.
Note that since no additional multi-looking is performed after phase linking, there are no wrapped phase misclosures in the re-formed interferograms; while this network is ``redundant'' in the wrapped domain, the phase unwrapping optimization problem often produces fewer errors and runs significantly faster for short-temporal-baseline interferograms. Network unwrapping is only necessary for decorrelating areas, as areas with high long-term coherence can perform a PS-like unwrapping strategy using all single-reference outputs.

Our algorithm has two options for phase unwrapping:

\begin{enumerate}
  \item Perform spatial unwrapping on each interferogram independently, using a 2D unwrapper such as SNAPHU \cite{Chen2001TwodimensionalPhaseUnwrapping}.
  \item Use the Extended Minimum Cost Flow (EMCF) approach \cite{Pepe2006ExtensionMinimumCost} to first unwrap in time, then apply the resulting $2\pi$ ambiguities to improve estimates of spatial phase gradients. Each interferogram is unwrapped in space independently \cite{Olsen2023ContextualUncertaintyAssessments}.
\end{enumerate}
In both cases, an unwrapped phase difference is obtained, from which a single-reference unwrapped phase time series at each SAR acquisition should be estimated via an inversion.

\subsection{Time series network inversion}
\label{subsec:methods-timeseries-inversion}

To set up the inversion, a spatial reference phase vector must be chosen, and the pixel's time series vector is subtracted from all pixels to remove arbitrary constants added by the unwrapping algorithm (Figure \ref{fig:processing-modules}e). To do this in an automated manner, we first select all pixels inside the largest unwrapping connected component with $\gamma_{t} > 0.95$ as reference pixel candidates. We then choose the pixel nearest to the centroid of this region, which ideally finds a pixel that has (1) low phase noise, and (2) is contained within a stable, contiguous area of high coherence, rather than being an outlier PS pixel at the frame boundary. Note that because no external calibration is used, the displacement results are internally referenced to the frame and do not represent absolute displacement.

With the spatial reference established, we now formulate the inversion problem. Because phase linking estimates a consistent phase vector, there is no wrapped phase misclosure when we re-form interferograms; the only errors that our inversion minimizes are integer unwrapping errors. 
Let $\mathbf{b} \in \mathbb{R}^{N_{\text{ifg}}}$ be the vector of measured unwrapped interferometric phase differences at a given pixel (one value per interferogram), and $\mathbf{x} \in \mathbb{R}^{N-1}$ be the phase time series relative to the first date that we wish to estimate. The unwrapping inversion problem can be written
\begin{equation}
	\mathbf{b} = A \mathbf{x} + 2\pi \mathbf{k}_{\text{error}},
\end{equation}
where $A \in \mathbb{R}^{N_{\text{ifg}} \times (N-1)}$ is the network incidence matrix, and $\mathbf{k}_{\text{error}} \in \mathbb{Z}^{N_{\text{ifg}}}$ is the vector of integer ambiguities resulting from unwrapping errors. If all pixels were unwrapped correctly, the vector $\mathbf{k}_{\text{error}}$ is all zero.

Rather than use least squares to estimate $\hat{\mathbf{x}}$ (which minimizes the $L_2$ norm of the residual $\mathbf{r} \isdef\mathbf{b} - A \hat{\mathbf{x}}$, \cite{Berardino2002NewAlgorithmSurface, Yunjun2019SmallBaselineInSAR}), we reduce the impact of phase unwrapping errors and promote a sparse residual vector by minimizing the $L_1$ norm \cite{Boyd2004ConvexOptimization}:
\begin{align}
  \underset{x}{\text{minimize}} &  \norm{A \mathbf{x} - \mathbf{b}}_1 .  \label{eq:l1-min}
\end{align}
While the $L_1$ minimization does not guarantee integer-multiple residuals of $2\pi$, the resulting solutions are often the same for pixels with a low percentage of unwrapping errors \cite{Lauknes2011InSARDeformationTime, Donoho2006CompressedSensing}.
Regions and/or dates with residuals significantly different from $2\pi k$ should be masked due to unwrapping errors.  We solve \eqref{eq:l1-min} using the Alternating Direction Method of Multipliers (ADMM) \cite{Boyd2010DistributedOptimizationStatistical}, which is particularly efficient for our problem structure and significantly reduces the computational cost compared to other integer-based solvers. See Appendix \ref{appendix:admm-details} for details on the ADMM implementation.

\subsection{Historical processing and incremental updates}
\label{subsec:methods-forward-mode}

Our algorithm can process either an existing historical archive, or ingest one new incremental image (i.e., forward processing) in near-real time. However, since total run time is dominated by phase unwrapping, unwrapping a full nearest-3 network of $3M - 6$ interferograms (where $M$ number of images in the mini-stack) can be inefficient.  To accelerate forward processing, a modified, smaller network of interferograms may be unwrapped. 

Unwrapping only the single new, nearest-neighbor interferogram accumulates errors when cumulatively summed \cite{Mirzaee2023NonlinearPhaseLinking} and using only short-temporal baseline interferograms may lead to a biased displacement estimate \cite{Ansari2021StudySystematicBias}. Therefore, we use all $K$ compressed SLCs and $M$ real SLCs to estimate an optimized wrapped phase, but we only unwrap the portion of the nearest-3 network which touches the newest acquisition. Compared to the full network of $39$ interferograms when $M=15$, unwrapping the $6$ interferograms which contain the four most recent dates runs in 10-20\% of the time.

Since we do not unwrap any interferograms containing the first acquisition in the mini-stack, we have two options for outputting new data:

\begin{enumerate}
	\item Output the latest displacement estimate relative to the previous date, which is included in the unwrapped network
	\item Fetch a previously-archived displacement containing the desired reference date, which has a secondary date in the smaller network, and add the displacement values from this previous product to all newly estimated displacement images.
\end{enumerate}

The first option simplifies processing, at the cost of adding burden on end users, while the second option increases the complexity of the data processing and archiving system.

\subsection{Summary of the Near-Real-Time Workflow}
To summarize, each time new SAR images arrive:
\begin{enumerate}
  \item Form a new mini-stack by adding the latest SLCs to the most recent compressed SLC.
  \item Update amplitude statistics (mean/variance) using \eqref{eq:merged-mean}--\eqref{eq:merged-variance}.
  \item Apply amplitude dispersion thresholds to pick PS pixels; use the GLRT to identify SHP neighborhoods for DS pixels.
  \item Perform phase linking on the sample coherence matrix \eqref{eq:sample-coherence}; form a new compressed SLC.
  \item Create the nearest-3 network of interferograms and perform phase unwrapping (full network of $3M-6$ interferograms for historical processing; smaller subset touching the newest date for forward processing; see Section~\ref{subsec:methods-forward-mode}).
  \item Invert the network using $L_1$ minimization to obtain the displacement time series at each acquisition date
  
\end{enumerate}

\section{Processing configuration for OPERA Displacement from Sentinel-1}\label{sec:processing-config}

Here we summarize the inputs and configuration choices used to produce the OPERA DISP-S1 products. We use the OPERA Level-2 Coregistered Single-Look Complex (CSLC-S1) products that are created by OPERA as an intermediate product required to produce OPERA DISP products \cite{opera_cslc_s1_atbd_2024,nasa_opera_2023}. The CSLC-S1 products are geocoded, single-burst images derived from the Level-1 radar-coordinate SLCs processed by the European Space Agency (ESA) Copernicus program. The topographic phase is removed using the Copernicus digital elevation model (DEM) \cite{ESA2020CopernicusDEM}. The geocoded SLCs are coregistered on a Universal Transverse Mercator (UTM) coordinate system at 5 meter by 10 meter posting in the X- and Y- directions, respectively. The spatial labeling of the CSLC-S1 used the Burst ID map provided by ESA \cite{SAR-MPCserviceSentinel1BurstID}. 

The OPERA CSLC-S1 and DISP-S1 products are produced for North America$^1$ geographic scope, defined as the U.S. and U.S. Territories, Canada within 200 km of the U.S. border, and all mainland countries from the southern U.S. border down to and including Panama. All available CSLC-S1 data between July 2016 -- today are used. There are notable differences in data collection between ascending and descending flight paths in North America$^1$ that result in significant differences in DISP-S1 data availability (Supplement Figure \ref{fig:supplement-opera-scope}).

Each DISP-S1 product was created on a fixed grid, nominally containing 27 bursts (9 bursts along-track for subswaths IW1, 2, and 3). All frames overlap by one burst along-track to aid in stitching displacement products. To ensure each Frame's displacement products have the same missing data pattern regardless of changes to the Sentinel-1 acquisition pattern, an analysis was done using the existing Sentinel-1 archive to discard off-nominal dates or Burst IDs (see Supplement \ref{sec:supplement-exclude-bursts} for details).

We used a mini-stack size of $M=15$ (approximately 6 months of data at 12-day repeat intervals) for all displacement processing, and we included the most recent 6 compressed SLCs in each mini-stack. For the operational production system, the compressed SLC reference is moved forward to the last real SLC of each mini-stack, producing short-temporal-baseline interferograms that are better suited for capturing rapid deformation events; see Supplement Section~\ref{sec:supplement-sequential-last-per-ministack} for a full discussion of this tradeoff. Our PS analysis (Section \ref{subsec:ps-shp-selection}) was done on the full-resolution geocoded SLCs.  Due to the small size of the input SLC stack, we set a conservative threshold of $0.2$ on the $D_A$ to select PS pixels, fixed across all geographic regions and processing batches.

For phase linking (Section \ref{subsec:proc-phase-linking}), our SHP window size varied depending on the land cover characteristics of each geographic region: for the higher coherence regions (e.g., the arid U.S.\ Southwest), we used a window of $25 \times 13$ pixels ($\sim 125 \times 130$ meters) to preserve more spatial detail, while for regions with dense vegetation (e.g., the Pacific Northwest and Alaska), where smaller windows led to clear unwrapping errors and poor GPS agreement, we increased the window up to $37 \times 19$ pixels ($\sim 185 \times 190$ meters). The larger windows trade spatial resolution for greater statistical reliability of the phase linking estimate, particularly in low-coherence environments. Complete window size configurations by region are listed in Supplement Section~\ref{sec:supplement-shp-window}. We targeted an output posting of $30 \times 30$ meters by performing the phase linking algorithm on every $6$th and $3$rd pixels in the X- and Y- directions, speeding up the phase linking step by a factor of $\sim$18. The output pixels used the phase of PS pixels (where available), and fell back to using the DS pixel phase otherwise (see Supplement \ref{sec:supplement-regridding} for details).

To maintain coherence in northern snow-prone regions, we systematically excluded winter acquisitions where the scattering surface has significantly changed from summer. We performed an analysis of the 2-meter temperature and snow cover variables from the NOAA Global Ensemble Forecast System (GEFS) reanalysis data, and we excluded SAR acquisitions during a ``blackout period'', which was determined on a per-frame basis. We found that including fewer scenes with higher quality led to significantly higher temporal coherence and fewer artifacts from unwrapping errors. See Supplement Section \ref{sec:supplement-snow} for complete methodology.

\section{Product Validation Approach}\label{sec:validation-methods}



Product ``Validation'' represents an independent and quantitative assessment of product quality, undertaken to verify that the retrieved measurements meet the accuracy, quality, and reliability standards. These standards are directly traceable to the project science objectives and product performance specifications established during the formulation and design phases. The OPERA Project has defined specific requirements for each of its products \cite{Bato2022OPERAValidationPlan}. For the OPERA DISP-S1 product, the requirement is as follows:

\begin{quote}
\textit{The Sentinel-1 DISP product (DISP-S1) shall measure surface displacement rates from a single line-of-sight geometry with an uncertainty of 5 mm per year or better over spatial scales between 0.1 km and 50 km, based on four years of regularly sampled Sentinel-1 A/B data. This requirement applies to at least 80\% of Sentinel-1 acquisitions with a 12-day repeat or better, using VV-polarization imagery in IW mode, in regions where the interferometric signal is maintained (i.e., coherence $>$ 0.5).
}
\end{quote}

\begin{figure}[!ht]
	\centering
	\includegraphics[width=0.99\columnwidth]{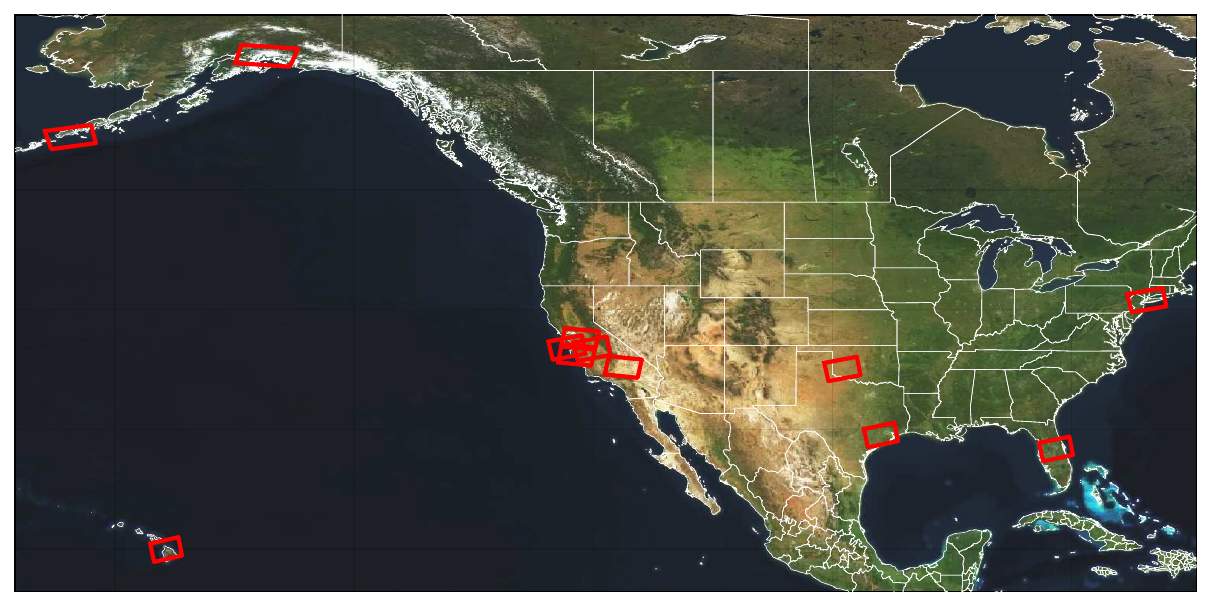}
	\caption{Calibration and Validation (CalVal) sites covering diverse thematic domains, including those related to volcanoes, tectonics/earthquakes, landslides, sinkholes, and land subsidence. These CalVal sites are selected to capture a range of environmental complexities and terrain types representative of the North America geographic scope.
	}\label{fig:validation_sites}
\end{figure}

Validation in the context of remote sensing, requires the collection of sufficient and independent ground-based reference data, and when available, complementary remotely sensed measurements to independently confirm the accuracy of the derived products. Given the OPERA DISP-S1 product's extensive spatial coverage over North America, our validation strategy ensures representative sampling across diverse terrains types, surface scattering, and environmental conditions. This comprehensive approach enables us to demonstrate product reliability and performance consistency across the North America geographic scope (Figure~\ref{fig:validation_sites}). 

The selected validation sites represent both negligibly deforming and deforming regions over North America (i.e., the latter is due to various natural and anthropogenic deformation processes such as vertical land motion, tectonics, and volcano deformation), with a range of seasonality and surface characteristics including, vegetation cover, rural, urban, or mixed environments, and terrain type (i.e., flat or mountainous). See Table S\ref{tab:disp_validation_summary} for details.

Building on the DISP-S1 requirements, we employ two complementary validation approaches to evaluate the DISP-S1 products: (1) Validation Approach 1 (VA1) relates to directly comparing DISP-S1 linear displacement rates and independent GPS velocity measurements projected into the radar look direction in regions with dense GPS coverage. (2) Validation Approach 2 (VA2) focuses on the InSAR residual analysis, which involves quantifying the velocity differences between random pixel pairs at different distances within 50~km, allowing assessment of the spatial correlation and noise characteristics. Because DISP-S1 is a frame-based product, the validation is likewise conducted on a frame-by-frame basis to ensure consistency with the product's spatial definition and processing framework. Below we elaborate the details of the two methods:

\subsection{Validation Approach 1 (VA1): Direct Comparison with GPS}
For each Frame~ID, the full DISP-S1 displacement time series is first reconstructed by connecting all mini-stacks and referencing each dataset to the earliest acquisition date. This reconstructed time series serves as the basis for deriving the line-of-sight (LOS) velocity estimates. In parallel, a temporal sampling ``pre-check'' is conducted to ensure that the required temporal coverage criteria are satisfied. 

Low-quality and unreliable pixels are removed using a reliability mask derived from the ``recommended mask'' layer included in each product. The recommended mask combines the water mask, temporal coherence ($<$~0.6), and phase similarity ($<$~0.5) thresholds to identify stable and coherent pixels. Only pixels that remain valid for at least 90\% of the acquisitions throughout the recommended mask time series are retained and evaluated for further analysis.

\begin{figure*}[!ht]
	\centering
	\includegraphics[width=0.95\textwidth]{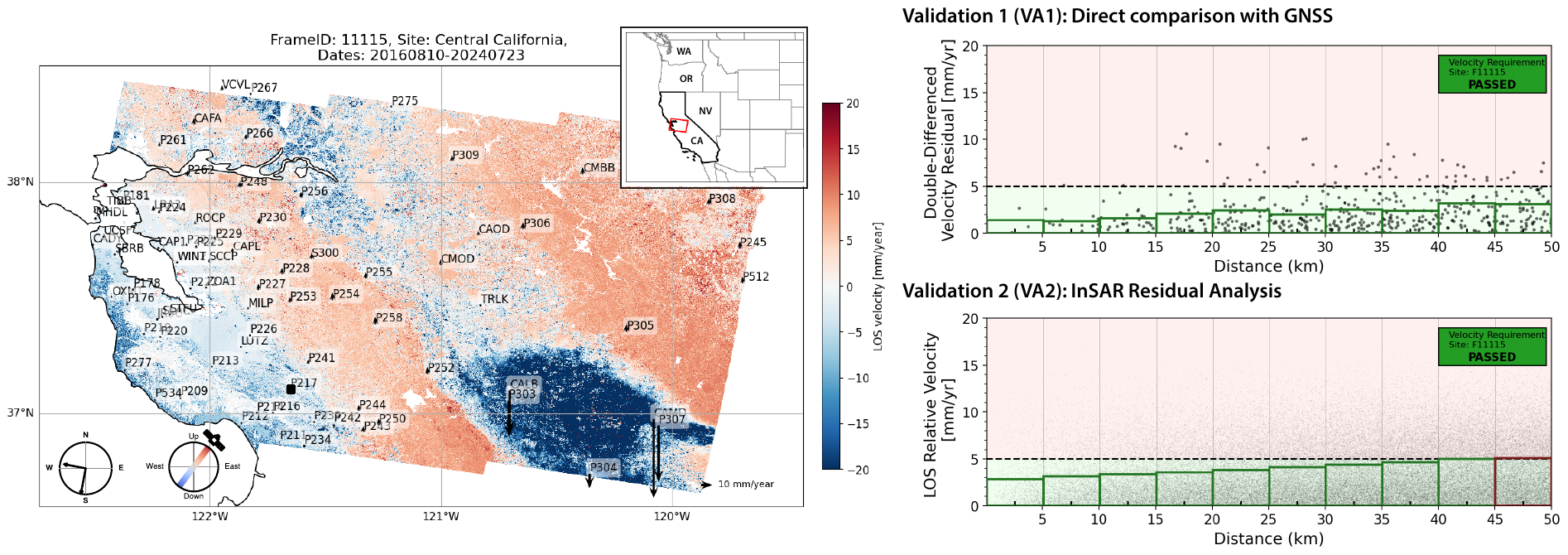}
	\caption{Example of validation analysis for a descending DISP-S1 frame (F11115) over Central California, USA. \textit{Left:} LOS velocity map derived from DISP-S1 products spanning October 2016 – 2024. \textit{Upper right:} Results of the double-differencing analysis between InSAR and cGPS measurements at 50 km spatial scales. \textit{Lower right:} Results of the InSAR residual analysis. For both VA1 and VA2, the dashed black line indicates the 5 mm/yr product accuracy requirement. Points above the line (pink region) exceed this threshold (fail), while points below the line (green region) meet the requirement (pass). Bars show $1~\sigma$ standard deviation for each distance-bin ($\sim5$~km). Green bars denote bins where residuals are within the 1$\sigma$ threshold (i.e., $\geq$68\% of points meet the requirement); red bars denote bins where residuals exceed it.}
	\label{fig:ex_validation_california}
\end{figure*}

GPS time series data are pre-processed before use in the DISP-S1 validation analysis. For each Frame~ID, GPS station data that are located within the frame boundaries are downloaded from publicly available sources (e.g., University of Nevada, Reno's GPS data products \cite{blewitt2018harnessing}) and then archived. The GPS time series are quality-checked and, where necessary, corrected for step anomalies such as those associated with antenna changes. The GPS LOS velocities are then computed over the same temporal span as the DISP-S1 velocity estimates to ensure temporal consistency between datasets. Similar to the NISAR validation approach for secular deformation, we form unique station pairs within 50 km and compare GPS and DISP-S1 LOS velocity double-differences for each pair \cite{chapman2024overview}. 
Assuming that the residual errors follow a Gaussian distribution, the DISP-S1 product is considered to meet the accuracy requirement if approximately 68\% of the residual values (i.e., within one standard deviation, $1\sigma$) fall below 5~mm/yr. In Figure~\ref{fig:ex_validation_california}, the bar plot colors indicate whether in each binned distance (i.e., roughly every 5 km), the residuals meet this criterion: green denotes “pass”, while red means “fail”. Each Frame~ID is ultimately assigned an overall \textbf{PASS} or \textbf{FAIL} designation based on these validation outcomes.

\subsection{Validation Approach 2 (VA2): InSAR Residual Analysis}
VA2 evaluates the noise characteristics over frames that in general cover negligible deformation or are within portions of the DISP-S1 frame that lie outside active deformation zones. For these ``negligibly deforming'' areas, we focus on pixels that maintain high coherence, as informed by the reliability mask.

We follow the same pre-processing steps as in VA1 to generate the DISP-S1 LOS velocity for each Frame~ID. In addition, regions exhibiting significant deformation within the frame are identified using z-score thresholding and subsequently masked out to isolate stable and coherent areas that are representative of the background noise characteristics. For each frame, pixels are randomly sampled within 50~km length scales and paired to compute LOS velocity differences. Similar to VA1, the residuals are assumed to follow a Gaussian distribution and that the requirement success threshold represents a 1$\sigma$ limit (i.e., 68\% of the residuals should fall under 5 mm/yr). The residual statistics for each binned distance are evaluated as in VA1 and each Frame~ID is assigned a \textbf{PASS} or \textbf{FAIL} designation based on compliance with these accuracy criteria.

\section{Results}\label{sec:results}

\begin{figure*}[!ht]
	\centering
	\includegraphics[width=0.99\textwidth]{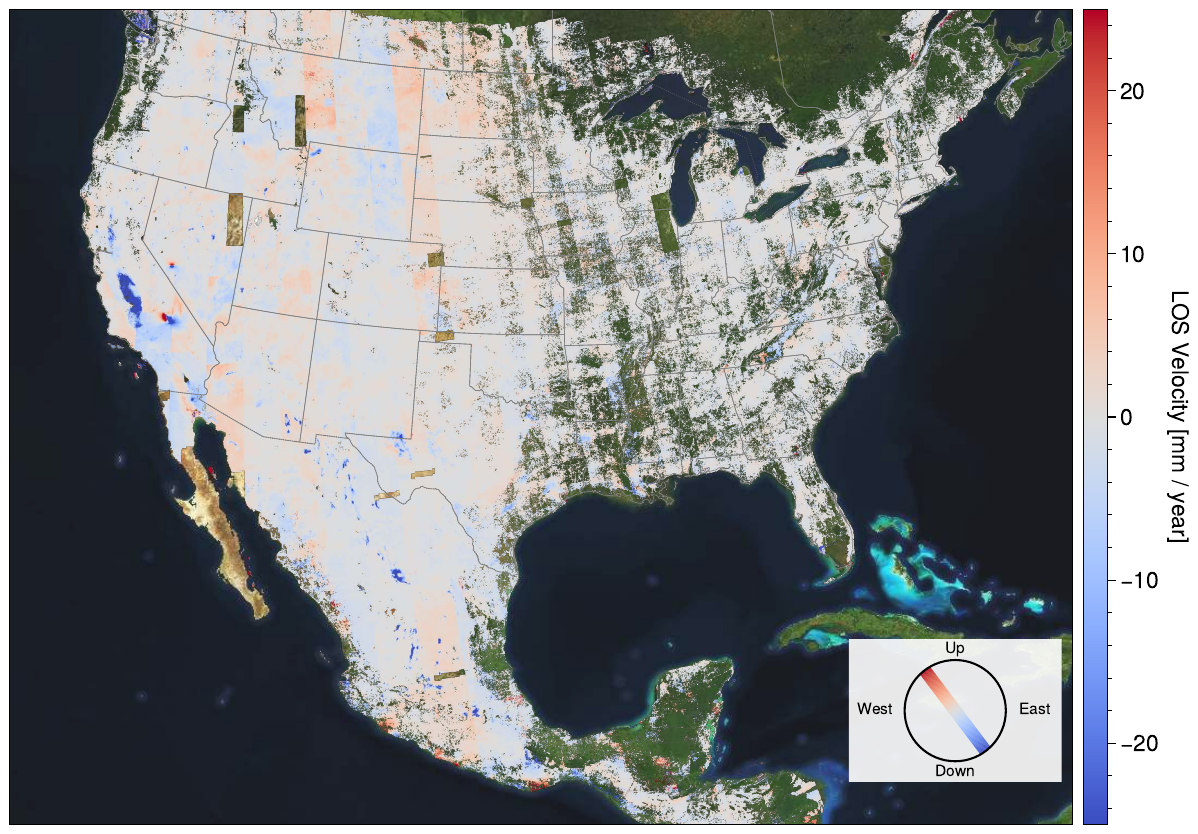}
	\caption{Average LOS surface velocity from Nov. 2017 to Nov. 2024 for the ascending Sentinel-1 geometry. Red indicates average motion toward the satellite, blue indicates motion away. Velocities are relative to an internal average rate per frame. For wide-area visualization continuity, a planar ramp was removed from each frame, and a constant shift was estimated for each frame using along-track overlapping regions. Descending LOS velocities over the same time period are shown in Supplement Figure \ref{fig:supplement-descending-conus-7year-velocity}.}
	\label{fig:results-conus-7year-velocity}
\end{figure*}

Our processing chain was used to produce over 170,000 displacement frames across North America$^1$. Figure \ref{fig:results-conus-7year-velocity} shows the average line-of-sight velocity for the ascending Sentinel-1 geometry for the region covering the continental U.S. and Central America. To create this velocity mosaic, a planar ramp was removed from each frame, and a constant offset per-frame was estimated using the median difference in the along-track overlaps. Note that no ionospheric or tropospheric corrections were applied to the data, and the velocities have not been calibrated to GPS \cite{Govorcin2024VariableVerticalLand}. We also note that ramp removal and frame offset estimation was performed only for visualization purposes; the OPERA DISP-S1 data products distributed to users retain the original unwrapped phases without ramp removal, preserving all long-wavelength signals including tectonic deformation and fault creep. The velocity mosaic captures a huge range of geophysical processes, including meter-scale subsidence in the Central Valley of California \cite{Govorcin2024VariableVerticalLand} and Mexico City \cite{Osmanoglu2011MexicoCitySubsidence}, from subsidence from agricultural groundwater withdrawal through the southwest U.S. and Mexico \cite{Cigna2021PresentdayLandSubsidence}, the 2019 Ridgecrest Earthquake \cite{Xu2020CoseismicDisplacementsSurface}, landslides in Los Angeles \cite{Handwerger2025MultisensorRemoteSensing, Li2024ExploringBehaviorsInitiated}, and the ongoing deformation in West Texas due to oil and gas production \cite{Hennings2023DevelopmentComplexPatterns,Staniewicz2020InSARRevealsComplex}. The descending LOS velocities over the same time period are shown in Supplement Figure \ref{fig:supplement-descending-conus-7year-velocity}, and quality metric summaries are shown in Supplement Figure \ref{fig:supplement-ascending-conus-quality}.

Below we examine two challenging case studies that demonstrate the method's performance: the Big Island of Hawaii, which includes 2018 Kilauea and 2022 Mauna Loa eruptions, and the Three Sisters volcanic complex in Oregon, which contains a subtle uplift signal hidden by strong vegetation and seasonal snow cover pose a serious challenge for C-band radar analysis.

\subsection{Kilauea, Hawaii}

\begin{figure}[!ht]
	\centering
	\includegraphics[width=0.99\columnwidth]{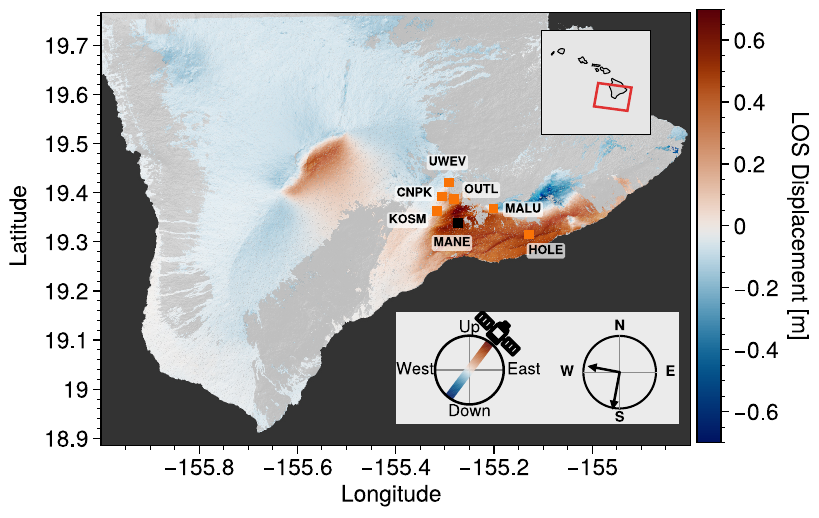}
	\caption{Cumulative LOS surface displacement (July 2016 - December 2024) for descending Track 87 (OPERA Frame 23211). Red indicates ground motion toward the satellite, blue indicates motion away. Orange squares indicate locations of GPS stations plotted in Figure \ref{fig:results-hawaii-gps-historical-comparison} (relative to the location shown as black square).
	}
	\label{fig:results-hawaii-eruption}
\end{figure}

\begin{figure}[!ht]
	\centering
	\includegraphics[width=0.99\columnwidth]{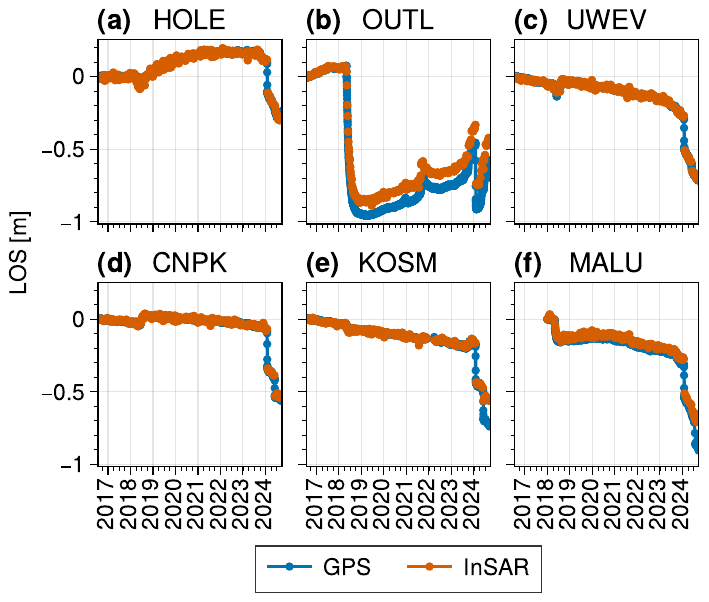}
	\caption{Comparison to cGPS stations (marked in Figure \ref{fig:results-hawaii-eruption}) projected along the descending radar line-of-sight direction). Blue points are GPS daily solutions, orange points are cumulative displacement values. The GPS station ``MANE'' was used as reference point for the GPS comparison}
	\label{fig:results-hawaii-gps-historical-comparison}
\end{figure}

\begin{figure}[!ht]
	\centering
	\includegraphics[width=0.99\columnwidth]{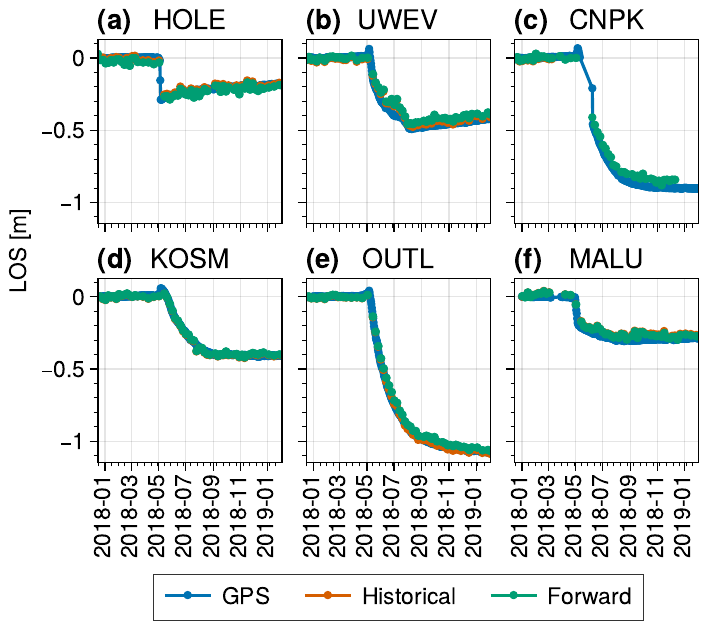}
	\caption{Comparison of ``Historical mode'' processing to ``Forward mode'' during the 2018 Kilauea eruption for ascending data. cGPS stations are projected along the radar line-of-sight direction. Blue points are GPS daily solutions, orange points are cumulative displacement values for Historical processing, green points are displacement values from Forward processing. The GPS station MANE was used as reference point for both Historical and Forward modes.}
	\label{fig:results-hawaii-forward-comparison}
\end{figure}

In 2018, Kilauea volcano produced its largest lower East Rift Zone eruption and summit collapse in over two centuries \cite{Neal20192018RiftEruption}, testing our algorithm's ability to handle rapid, large-magnitude deformation in an automated processing framework. The event produced cumulative displacements exceeding 1 meter in the rift zone and summit region (Figure~\ref{fig:results-hawaii-eruption}), with deformation rates during the main collapse period exceeding centimeters per day.

Figure~\ref{fig:results-hawaii-eruption} shows cumulative line-of-sight displacement from July 2016 to December 2024 for descending Track 87 (OPERA Frame 23211). The spatial pattern resolves both the summit collapse and the lower East Rift Zone inflation/deflation cycle. The dataset also captured the November 2022 Mauna Loa eruption, including the subtle pre-eruptive inflation beginning in 2021 and the accelerated deformation in September 2022 \cite{Lynn2024Triggering2022Eruption, Ellis2024DeformationMaunaLoa}. Despite the extreme deformation magnitudes and rates, our sequential phase linking approach with L1-norm network inversion successfully maintains displacement continuity across both events without manual intervention.

To validate accuracy, we compared our historically-processed InSAR-derived displacements to continuous GPS (cGPS) stations spanning the deformation field (Figure~\ref{fig:results-hawaii-gps-historical-comparison}). GPS positions were projected into the radar line-of-sight direction and referenced to station MANE. The time series show excellent agreement throughout the observation period, including during the rapid collapse phase in mid-2018. Root mean square differences between DISP-S1 and GPS range from 0.9 cm to 3.4 cm across five stations with high coherence; station OUTL, closest to the caldera collapse and experiencing the lowest coherence, shows 9.4 cm RMSE.

To verify that our Forward mode processing (Section \ref{subsec:methods-forward-mode}) produces results consistent with Historical mode, we reprocessed 60 acquisitions from late 2017 to early 2019 spanning the Kilauea eruption using both approaches. We observe similar accuracy in our incremental, NRT Forward mode as the Historical mode processing when validated against GPS (Figure~\ref{fig:results-hawaii-forward-comparison}).

\subsection{Three Sisters, Oregon}

\begin{figure}[!ht]
	\centering
	\includegraphics[width=0.99\columnwidth]{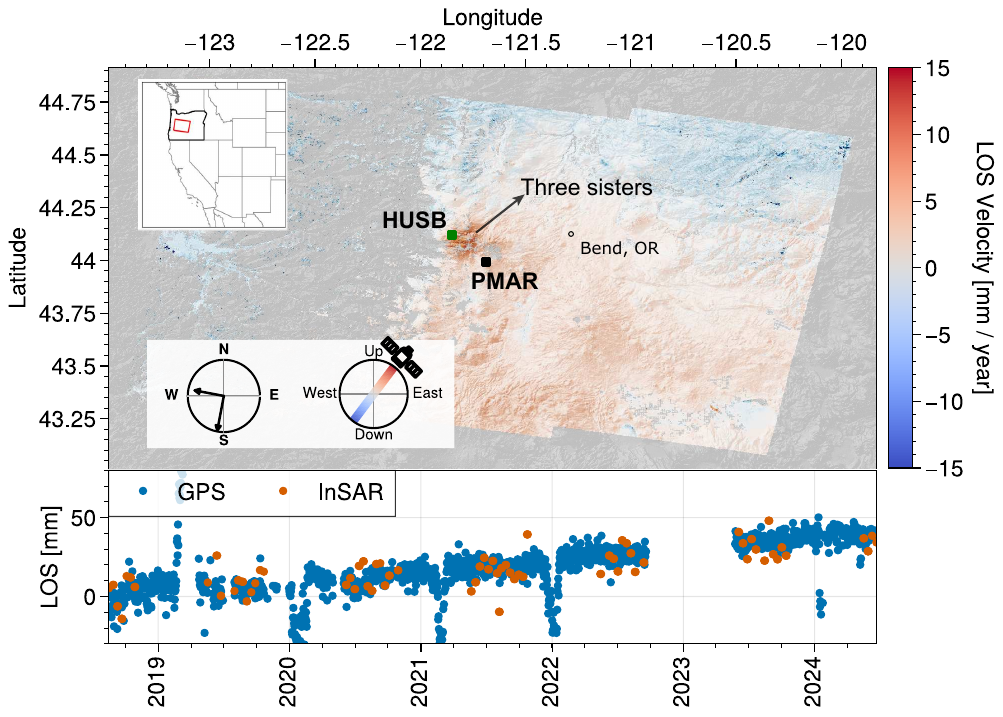}
	\caption{Cumulative line-of-sight displacement at Three Sisters volcanic complex (November 2017-December 2024, Descending Track 115, OPERA Frame 30710). Upper: Spatial deformation pattern showing peak uplift west of South Sister. Green marker is the location of GPS station HUSB, black marker is reference GPS stations PMAR.
	Lower: Time series comparison between DISP-S1 (orange) and GPS at station HUSB (blue), referenced to station PMAR.}
	\label{fig:results-oregon-uplift}
\end{figure}

The Three Sisters volcanic complex in central Oregon has exhibited episodic uplift since the mid-1990s, attributed to magma intrusion at approximately 6 km depth \cite{Lisowski2021GeodeticConstraints25year}. Following the initial rapid inflation period (1998-2000) that produced ~30 cm of uplift, the deformation rate has declined exponentially, with recent rates of approximately 3 mm/yr. Detecting this subtle ongoing signal with C-band InSAR is particularly challenging due to dense coniferous forest cover and seasonal snow accumulation that degrades coherence throughout much of the year.

Figure \ref{fig:results-oregon-uplift} shows cumulative line-of-sight displacement from September 2017 to December 2024 for descending Track 115 (OPERA Frame 30710). The spatial pattern of deformation is clearly resolved, with peak uplift centered approximately 5 km west of South Sister, consistent with the known source location. Despite the challenging environmental conditions, our NRT sequential phase linking approach maintains sufficient coherent pixel density to map the deformation field.

The time series comparison at GPS station HUSB (referenced to station PMAR) demonstrates excellent agreement between InSAR and GPS measurements throughout the observation period (Figure \ref{fig:results-oregon-uplift}, lower panel). The InSAR solution successfully tracks the subtle linear uplift trend with millimeter-level precision.

We note that for this frame, including winter/spring scenes with snow cover led to significantly lower temporal coherence and ambiguous signals in the uplifting area. This temporal filtering strategy, by retaining only the most coherent observations, enabled reliable volcanic deformation monitoring in an environment traditionally considered challenging for C-band InSAR analysis.

\subsection{Product Validation}

In Table \ref{tab:disp_validation_summary}, we summarized the DISP-S1 validation results. Results show that the DISP-S1 product meets and exceeds its defined mission accuracy requirements. The DISP-S1 requirement specifies that the products must achieve the displacement rate accuracy threshold over at least 80\% of all the validation sites evaluated. Based on the combined outcomes of VA1 and VA2, DISP-S1 successfully meets the product requirement criterion at over 96\% of all validation sites analyzed. Note that Frames F42779 (Mainland Alaska), F33065 (Unimak Island), F12640 (Florida), and F28486 (Oklahoma) were excluded from the VA1 analysis due to sparse cGPS network coverage. The California frame F11116 was also excluded from the VA2 analysis because a large portion of the frame exhibits significant deformation, making it unsuitable for background noise assessment.

For transparency, reproducibility, and traceability of the validation process, OPERA publicly releases the ``DISP-S1 Calibration and Validation (Cal/Val) Database''. The database includes all cGPS datasets used in the evaluation across all validation sites, along with associated metadata for DISP-S1 Cal/Val products and links to their corresponding online archive locations. It also contains a reference document describing the database structure, contents, and usage, as well as a summary report presenting the DISP-S1 validation results. All relevant validation information, methods, and reference data have been archived at the ASF DAAC at \url{https://www.earthdata.nasa.gov/data/projects/opera}.

\begin{table*}[!htb]
    \caption{Summary of the product validation results from the VA1 (direct comparison with cGPS) and VA2 (InSAR residual analysis) approaches for each validation site.
    }
    \centering
    \small
    \renewcommand{\arraystretch}{1.2}

    \begin{tabular}{llccc}
        \toprule
        \textbf{Frame ID} & 
        \textbf{Date Range} & 
        \textbf{\# DISP-S1 Data} & 
        \textbf{VA1} & 
        \textbf{VA2} \\
        \hline
        F11115 (California) & 20160810--20240723 & 299 & Passed & Passed\\ 
        F11116 (California) & 20160705--20241003 & 314 & Passed & -- \\ 
        F08882 (Houston) & 20160927--20240804 & 224 & Passed & Passed \\ 
        F42779 (Alaska) & 20180808--20241128 & 44 & -- & Passed \\ 
        F33065 (Unimak Island) & 20160723--20210925 & 59 & -- & Failed \\ 
        F36542 (California) & 20160724--20240730 & 329 & Passed & Passed \\ 
        F12640 (Florida) & 20160705--20241214 & 224 & -- & Passed \\ 
        F18903 (Ridgecrest) & 20160719--20230625 & 269 & Passed & Passed \\ 
        F08622 (New York) & 20160716--20241213 & 239 & Passed & Passed \\ 
        F09156 (California) & 20160705--20241003 & 299 & Passed & Passed \\ 
        F28486 (Oklahoma) & 20160710--20240926 & 224 & -- & Passed \\ 
        F33039 (Hawaii) & 20181017--20221009 & 210 & Passed & Passed \\
        \\
        \textbf{Summary} &  &  & 100\% Passed & 91\% Passed \\ 
        \bottomrule
        \end{tabular}
    \label{tab:disp_validation_summary}
\end{table*}

\section{Discussion}

\subsection{Computational efficiency and run time estimates}
\label{sec:discussion-runtimes}

\begin{figure}[!ht]
	\centering
	
	\includegraphics[width=0.95\columnwidth]{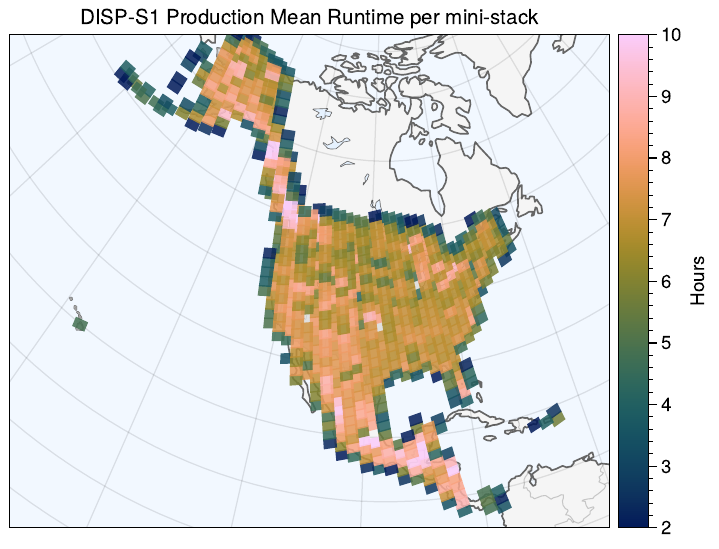}
	\caption{Average run time (in hours) for each mini-stack by geographic location. Frames near land borders have fewer than a full 27-burst frame.}
	\label{fig:discussion-runtime}
\end{figure}

\begin{table}

	\caption{Summary of run times (in hours) of each workflow for $n = 6062$ processed mini-stacks}
	\label{table:runtimes}
	\begin{center}
\begin{tabular}{lrrrr}
	\toprule
	 & Mean & Median & 10th \% & 90th \%  \\
	\midrule
	Total  & 6.70 & 6.70 & 5.49 & 7.99 \\
	Phase unwrapping & 5.39 & 5.34 & 4.29 & 6.61 \\
	\bottomrule

\end{tabular}
\end{center}
\end{table}
We evaluated the run time performance for the DISP-S1 workflow across the full North America production (over 6,000 mini-stacks of 15 dates each). Figure~\ref{fig:discussion-runtime} shows the mean wall-clock time per mini-stack by geographic location. Typical values are 6--8~hours, with longer run times occurring in regions with more challenging phase unwrapping (e.g., Alaska and Central America). Note that the shorter run times along the geographic boundary occur where frames contain fewer than the full 27 burst IDs. All runs were executed on Amazon Elastic Compute Cloud (EC2) instances provisioned with 32~vCPUs and 64~GiB of memory. We used the following instance families: \texttt{c5a.4xlarge}, \texttt{c6a.4xlarge}, \texttt{c6i.4xlarge}, \texttt{c7a.4xlarge}, and \texttt{c7i.4xlarge}.

A statistical summary of run times is given in Table~\ref{table:runtimes}, showing that phase unwrapping dominates the total Historical processing time. Median run times were 6.7~hours per mini-stack in total, of which $\sim$80\% is associated with phase unwrapping. Note that the wrapped phase estimation (phase linking, persistent scatterer selection) has a very narrow spread and consistently runs in $\sim$90 minutes per mini-stack.
Additionally, processing in Forward mode leads to a drop of 80-90\% in unwrapping run time, or about a 50-60\% reduction in total run time.

\subsection{Effect of mini-stack size}
\label{sec:discussion-ministack-size}

\begin{figure}[!ht]
	\centering
	\includegraphics[width=0.99\columnwidth]{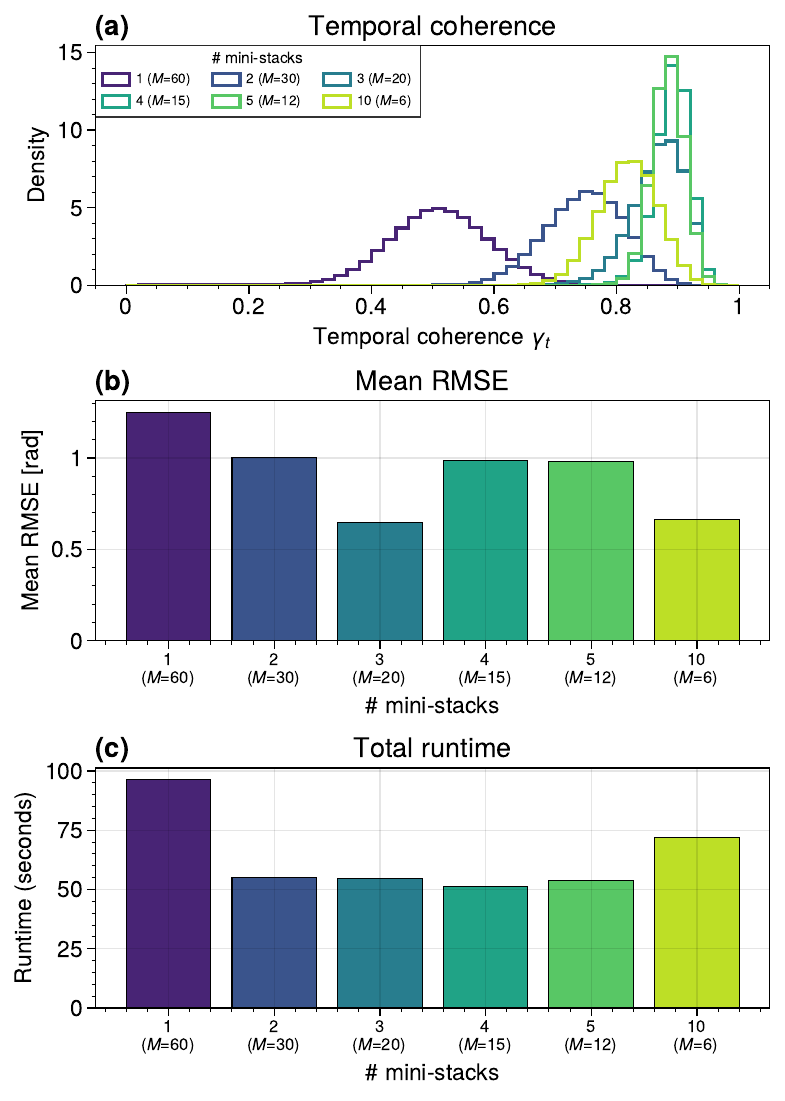}
	\caption{Effect of mini-stack size on estimation quality and run time for the synthetic dataset ($N = 60$ SLCs, $\tau = 60$~days).
		(a)~Distribution of average temporal coherence $\gamma_t$ for different mini-stack sizes. $\gamma_t$ has limited dynamic range for small $M$, inflating the apparent quality.
		(b)~Mean RMSE across all dates, showing that the full-stack case ($M = 60$) has the highest error while splitting into mini-stacks yields lower, relatively stable RMSE.
		(c)~Total phase linking run time as a function of the number of mini-stacks, showing a U-shaped curve due to the trade-off between covariance matrix size and I/O overhead.}
	\label{fig:ministack-size-comparison}
\end{figure}

We investigated the effect of the choice of $M$, the mini-stack size, on both quality of the phase-linked estimates and the computational cost. We used the same synthetic dataset described in Section~\ref{subsec:synthetic-ifg-demo} ($N=60$ SLCs, $\tau = 60$~days), varying the number of mini-stacks from 1 (i.e., batch processing the full stack with $M = 60$) to 10 mini-stacks of $M = 6$ images (Figure~\ref{fig:ministack-size-comparison}). The run time (Figure~\ref{fig:ministack-size-comparison}c) follows a U-shaped curve: a single large mini-stack is the most expensive due to the $O(M^2)$ growth of the sample coherence matrix, while very small mini-stacks incur overhead from repeatedly loading and writing data for overlapping spatial locations and from accumulating compressed SLCs in each batch. The minimum run time occurs at intermediate sizes ($M \approx 12$--$25$).

The mean RMSE (Figure~\ref{fig:ministack-size-comparison}b) shows a degradation of estimation quality for large ($M>50$) mini-stacks. The single-batch case ($M = 60$) produces the highest RMSE, while splitting into 2 or more mini-stacks yields lower error. This effect was noted by \cite{Ansari2017SequentialEstimatorEfficient} and \cite{Mirzaee2023NonlinearPhaseLinking}, who observed that batching the input SLC stack into mini-stacks can outperform the full-stack MLE in the presence of fast decorrelation. Large mini-stacks include long-temporal-baseline pairs that are fully decorrelated, yet whose sample coherence magnitudes are biased upward from zero due to the finite-sample bias of the coherence estimator \cite{Touzi1999CoherenceEstimationSAR}. Partitioning the stack into mini-stacks naturally limits the temporal baselines considered, avoiding the incorporation of noise-dominated pairs. For mini-stack sizes between $M = 6$ and $M = 30$, the mean RMSE is relatively stable, indicating that quality is not highly sensitive to the exact choice of $M$ within this range.

The temporal coherence distributions (Figure~\ref{fig:ministack-size-comparison}a) illustrate a complementary point. While $\gamma_t$ appears to improve monotonically as mini-stacks get smaller, this trend is partly an artifact: the $\gamma_t$ metric (Eq.~\ref{eq:temporal-coherence}) has limited dynamic range for small $M$ (for $M = 2$, $\gamma_t = 1$ identically; see Section~\ref{subsec:methods-similarity}). The mean RMSE (panel~(b)) shows that the actual estimation quality does not directly correlate with higher $\gamma_t$. This ceiling effect is one motivation for the complementary phase similarity metric $\gamma_s$, as well as for choosing a mid-sized value of $M$. For the OPERA DISP-S1 production, we use $M = 15$ with up to $K = 6$ compressed SLCs (Section~\ref{sec:processing-config}), which balances computational efficiency with estimation quality. One final note is that while the mini-stack size need not be identical across batches, since smaller mini-stacks lead to higher $\gamma_t$ values, a change in mini-stack size makes direct comparison of optimization quality more challenging for differently-sized mini-stacks.

\section{Conclusion}
We presented a comprehensive surface displacement processing chain capable of near-real-time updates within hours of new SAR acquisitions without historical archive reprocessing. The sequential phase linking approach estimates an optimized wrapped phase from compact ``mini-stacks'' of coregistered SLCs, enabling reliable deformation retrieval even under challenging coherence conditions. The compressed SLC architecture provides a practical foundation for scalable, cloud-based deployment.

Using results from the $\sim$170,000 displacement products generated for the OPERA Surface Displacement from Sentinel-1 dataset over North America, the algorithm demonstrates robust performance and strong agreement with GPS, with mean velocity differences under 4 $\mathrm{mm\,yr^{-1}}$ at length scales under 50 kilometers. Our approach captures signals ranging from meter-scale co-eruptive displacement at Kilauea to millimeter-per-year uplift signals in snowy, heavily vegetated regions. The computational efficiency of our open-source libraries has significantly lowered the barrier for large-scale processing of SAR datasets for both science and application use cases.

\appendices
\section{Details on ADMM Algorithm for \texorpdfstring{$L_1$}{L1} Minimization}
\label{appendix:admm-details}

The ADMM algorithm rewrites the $L_1$ minimization problem as \cite{Boyd2010DistributedOptimizationStatistical}
\begin{align*}
  & \underset{\mathbf{x},\mathbf{z},\mathbf{u}}{\text{minimize}} & & \lVert \mathbf{z} \rVert_{1} \\
  & \text{subject to} & & A\mathbf{x} - \mathbf{z} = \mathbf{b}
\end{align*}

where $\mathbf{u}$ is a dual variable.
At iteration $k$,

\begin{align}
  \mathbf{x}^{k+1} &= (A^T A)^{-1}A^T(\mathbf{b} + \mathbf{z}^k - \mathbf{u}^k) \\
  \mathbf{z}^{k+1} &= S_{1/\rho}(A\mathbf{x}^{k+1} - \mathbf{b} + \mathbf{u}^k) \\
  \mathbf{u}^{k+1} &= \mathbf{u}^k + A\mathbf{x}^{k+1}  - \mathbf{b} - \mathbf{z}^{k+1}
\end{align}
with $\rho > 0$ an augmented Lagrangian parameter, and $S_{\kappa}(\cdot)$ the soft-thresholding operator:
\begin{equation}
  S_{\kappa}(a) =
  \begin{cases}
    a - \kappa & a > \kappa  \\
    0 & |a| < \kappa \\
    a + \kappa & a < -\kappa
  \end{cases}.
\end{equation}
Since $A$ is the same for all pixels, we can precompute the Cholesky factorization of $A^T A$ once for all pixels.

\section*{Acknowledgments}

The authors would like to thank Zhong Lu and Steven Chan for their guidance to the OPERA project validation team, Luca Cinquini, Philip Yoon and Scott Collins for their work on the OPERA Science Data System, and Virginia Brancato, Seongsu Jeong and Liang Yu for their work creating a high-quality CSLC-S1 product.

The OPERA Level-2 CSLC-S1 and OPERA Level-3 Displacement from Sentinel-1 products are free and openly available at the ASF DAAC.
The DEM used during processing was modified to be a combination of the Copernicus DEM 30-m and Copernicus DEM 90-m models provided by the European Space Agency. The Copernicus DEM 30-m and Copernicus DEM 90-m were produced using Copernicus WorldDEM-30 © DLR e.V. 2010-2014 and © Airbus Defence and Space GmbH 2014-2018 provided under COPERNICUS by the European Union and ESA; all rights reserved. NOAA Global Ensemble Forecast System (GEFS) data, reprocessed by dynamical.org, was accessed on August 3, 2025 at \url{https://data.dynamical.org/noaa/gefs/analysis/latest.zarr}.

OPERA, managed by the Jet Propulsion Laboratory and funded by the Satellite Needs Working Group, is creating remote sensing products to address Earth observation needs across U.S. civilian federal agencies.
This research was carried out at the Jet Propulsion Laboratory, California Institute of Technology, under a contract with the National Aeronautics and Space Administration (80NM0018D0004).

\ifCLASSOPTIONcaptionsoff
\newpage
\fi


\bibliographystyle{IEEEtran}
\bibliography{IEEEabrv,citations}

\begin{thebibliography}{1}
\providecommand{\url}[1]{#1}
\csname url@samestyle\endcsname
\providecommand{\newblock}{\relax}
\providecommand{\bibinfo}[2]{#2}
\providecommand{\BIBentrySTDinterwordspacing}{\spaceskip=0pt\relax}
\providecommand{\BIBentryALTinterwordstretchfactor}{4}
\providecommand{\BIBentryALTinterwordspacing}{\spaceskip=\fontdimen2\font plus
\BIBentryALTinterwordstretchfactor\fontdimen3\font minus
  \fontdimen4\font\relax}
\providecommand{\BIBforeignlanguage}[2]{{%
\expandafter\ifx\csname l@#1\endcsname\relax
\typeout{** WARNING: IEEEtran.bst: No hyphenation pattern has been}%
\typeout{** loaded for the language `#1'. Using the pattern for}%
\typeout{** the default language instead.}%
\else
\language=\csname l@#1\endcsname
\fi
#2}}
\providecommand{\BIBdecl}{\relax}
\BIBdecl

\bibitem{ewert20182018}
D.~A. K. R. D.~W. Ewert, John~W., ``2018 update to the {{US Geological Survey}}
  national volcanic threat assessment,'' US Geological Survey, Tech. Rep.,
  2018.

\end{thebibliography}


\begin{IEEEbiography}[{%
	\includegraphics[width=1in,height=1.25in,clip,keepaspectratio]{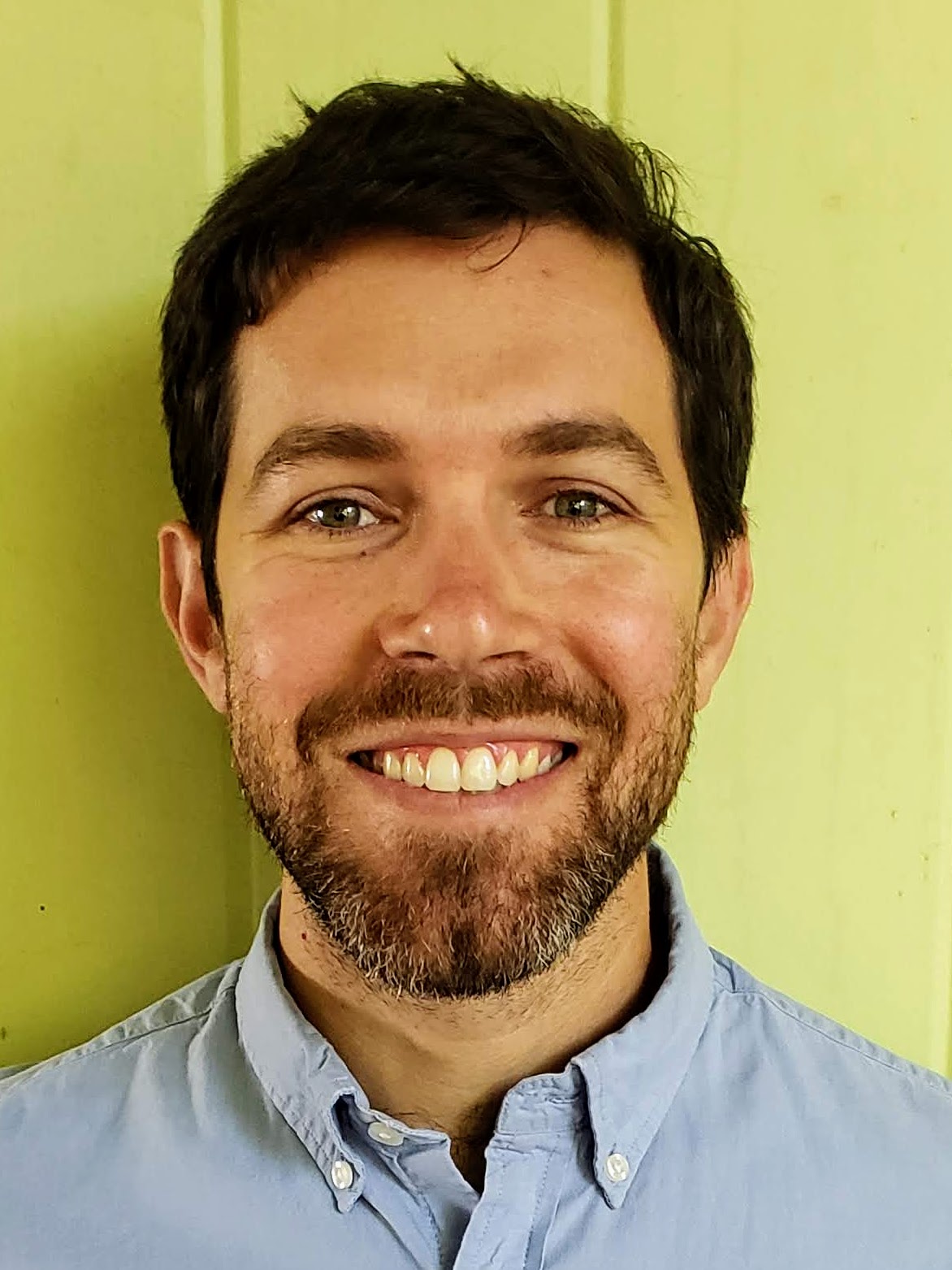}
	}]{Scott Staniewicz}
	received the B.S. degree in Electrical Engineering and Mathematics from Tufts University, Medford, MA, USA, in 2013, and the Ph.D. degree in Aerospace Engineering from the University of Texas at Austin, Austin, TX, USA, in 2022. He is currently a Signal Analysis Engineer at the Jet Propulsion Laboratory, California Institute of Technology. He is the Surface Displacement Processing (DISP) Product Lead for the OPERA project. His research interests include large-scale interferometric synthetic aperture radar (InSAR) processing and time-series analysis, statistical estimation, uncertainty quantification, and computer vision.
\end{IEEEbiography}

\begin{IEEEbiography}[{%
	\includegraphics[width=1in,height=1.25in,clip,keepaspectratio]{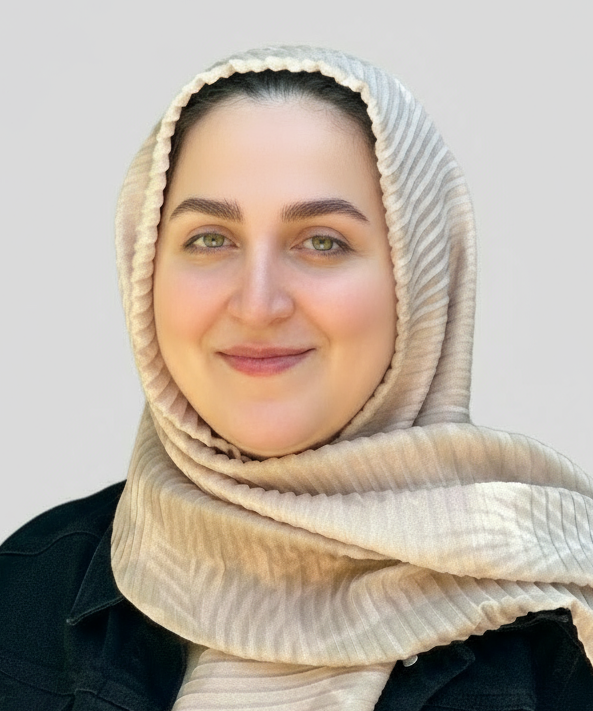}
	}]{Sara Mirzaee}
	received the B.Eng. degree in Surveying and Geomatics Engineering from the University of Zanjan, Zanjan, Iran, in 2012, the M.S. degree in Remote Sensing from the University of Tehran, Tehran, Iran, in 2015, and the Ph.D. degree in Geological and Earth Sciences from the University of Miami, Miami, FL, USA, in 2022. From 2022 to 2023, she was a Postdoctoral Scholar at the California Institute of Technology. She is currently a Signal Analysis Engineer at the Jet Propulsion Laboratory, California Institute of Technology, Pasadena, CA, USA, where she develops algorithms for near real-time ground displacement time-series processing for the OPERA project.   Her research interests include large-scale interferometric synthetic aperture radar (InSAR) processing, phase linking, time-series analysis, and algorithm development for operational Earth surface monitoring.
\end{IEEEbiography}

\begin{IEEEbiography}[{%
	\includegraphics[width=1in,height=1.25in,clip,keepaspectratio]{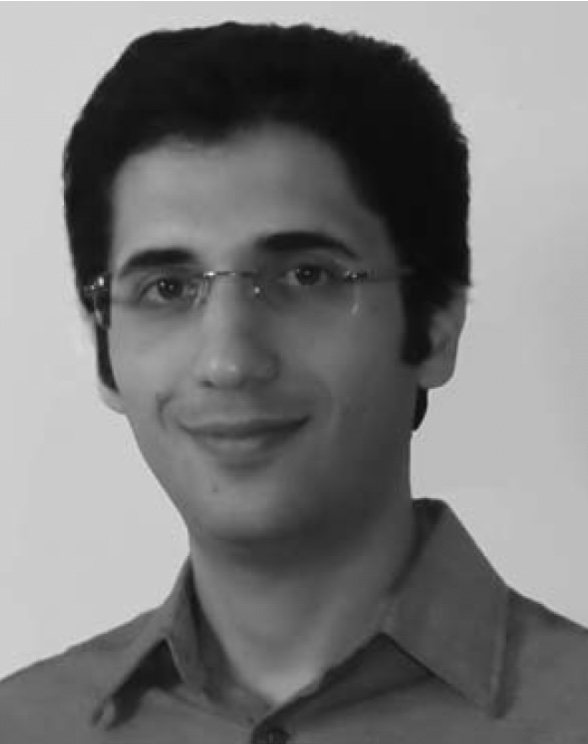}%
	}]{Heresh Fattahi}
	(M'12) received the M.S. degree in remote sensing engineering from the K. N. Toosi University of Technology, Tehran, Iran, in 2007, and the Ph.D. degree in geosciences from the University of Miami, Coral Gables, FL, USA, in 2015.,He was a Postdoctoral Scholar at the California Institute of Technology from 2015 until June 2017. Since then he has joined the Radar Algorithms and Processing Group, Jet Propulsion Laboratory, Pasadena, CA, USA. His research interests include algorithm development for SAR, InSAR, and InSAR time-series analysis.
\end{IEEEbiography}

\begin{IEEEbiography}[{%
	\includegraphics[width=1in,height=1.25in,clip,keepaspectratio]{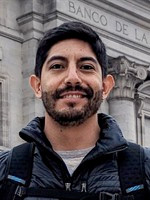}
	}]{Talib Oliver-Cabrera}
    received his B.S. degree in Geomatics Engineering from the National Autonomous University of Mexico, Mexico City, in 2010, and his Ph.D. degree in Marine Geology and Geophysics from the University of Miami, Miami, FL, in 2018. He is currently a Signal Analysis Engineer at the Jet Propulsion Laboratory (JPL), California Institute of Technology, Pasadena, CA, where he develops SAR and InSAR processing algorithms for the Sentinel-1 Disturbance and Vertical Land Motion products within the OPERA project. His research interests include wetland monitoring, Earth surface deformation, sea ice motion, natural hazard monitoring, and satellite and airborne SAR/InSAR applications.
\end{IEEEbiography}

\begin{IEEEbiography}[{%
	\includegraphics[width=1in,height=1.25in,clip,keepaspectratio]{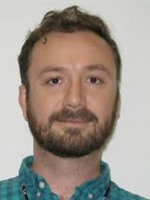}%
	}]{Emre Havazli}
    received the B.S. degree in Geodesy and Photogrammetry Engineering from Yildiz Technical University, Istanbul, Türkiye, in 2009, the M.S. degree in Geodesy from Boğaziçi University, Istanbul, Türkiye, in 2012, and the Ph.D. degree in Marine Geology and Geophysics from the University of Miami, Coral Gables, FL, USA, in 2018. He is a Signal Analysis Engineer with the Jet Propulsion Laboratory, California Institute of Technology, Pasadena, CA, USA, where he develops SAR and InSAR algorithms for the NISAR Solid Earth Science Team. His research interests include crustal deformation, snow water equivalent estimation, and disaster monitoring and response using satellite and airborne SAR/InSAR.
\end{IEEEbiography}

\begin{IEEEbiography}[{%
    \includegraphics[width=1in,height=1.25in,clip,keepaspectratio]{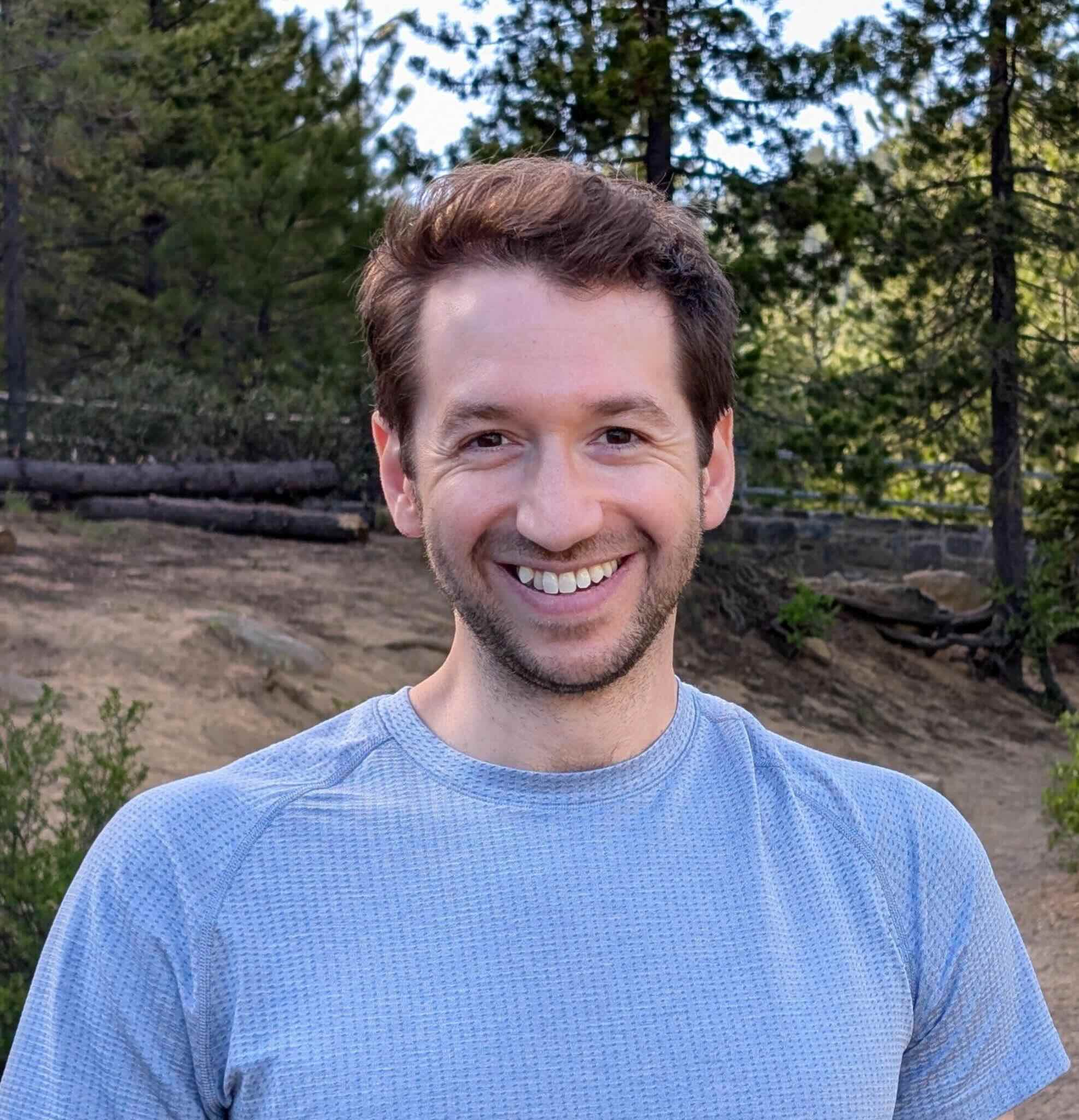}%
	}]{Geoffrey Gunter}
	received his B.S. degree in Biomedical Engineering from Washington University in St. Louis, MO, USA, in 2012, and his M.S. degree in Biomedical Engineering from Stony Brook University, NY, USA, in 2018. He worked as a Signal Analysis Engineer at the Jet Propulsion Laboratory, California Institute of Technology from 2018 until 2025, where he was a member of algorithm development teams for the NISAR mission and the OPERA project. He is currently a Radar Algorithms Engineer at Array Labs in Palo Alto, CA, USA. His research interests include multi-static radar, tomographic applications of SAR, phase unwrapping, and parallel and distributed computing.
\end{IEEEbiography}

\begin{IEEEbiography}[{%
		   \includegraphics[width=1in,height=1.25in,clip,keepaspectratio]{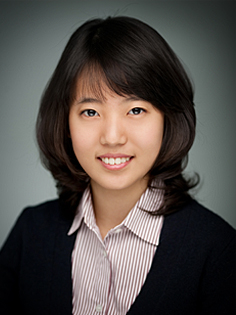}%
	}]{Se-Yeon Jeon}
	received the B.S. and Ph.D. degrees in IT Engineering from Yonsei University, Seoul, South Korea, in 2014 and 2019, respectively. From 2019 to 2021, she was a Postdoctoral Researcher with the German Aerospace Center (DLR), Oberpfaffenhofen, Germany, where she worked on waveform-encoded synthetic aperture radar (SAR) and NewSpace SAR concepts. She subsequently worked with Hanwha Systems Co. and the Agency for Defense Development (ADD), South Korea, on defense-oriented SAR system development. Since 2022, she has been a Signal Analysis Engineer with the Jet Propulsion Laboratory (JPL), California Institute of Technology, Pasadena, CA, USA, where she has worked on distributed drone-borne SAR experiments and algorithms. Her research interests include advanced SAR concepts and system design, and signal processing.
\end{IEEEbiography}

\begin{IEEEbiography}[{%
   \includegraphics[width=1in,height=1.25in,clip,keepaspectratio]{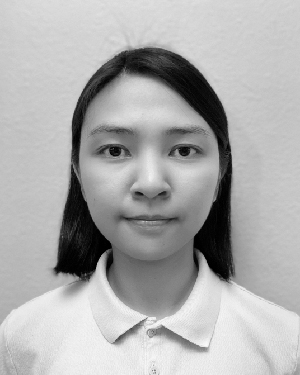}%
	}]{Mary Grace Bato} received the B.S. degree in Applied Physics from the University of the Philippines-Diliman, Philippines, in 2008, the M.S. degree in Earth Sciences from Université Blaise Pascal, Clermont-Ferrand II, Clermont-Ferrand, France, in 2012, and the Ph.D. degree in Earth, Universe, and Environmental Sciences from Université Grenoble Alpes, Grenoble, France, in 2018. She is currently a Signal Analysis Engineer at the Jet Propulsion Laboratory, California Institute of Technology, where she serves as the Calibration and Validation Lead for the OPERA Coregistered Single Look Complex (CSLC) and Displacement (DISP) products. Her research interests include understanding crustal deformation processes using remote sensing datasets, developing model–data fusion techniques, and transforming large-scale satellite data into actionable information that bridges the gap between scientific research and disaster response, particularly in volcano forecasting and hazard management.
\end{IEEEbiography}

\begin{IEEEbiography}[{%
	   \includegraphics[width=1in,height=1.25in,clip,keepaspectratio]{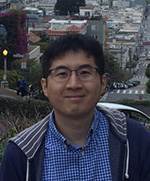}%
	}]{Jin-Woo Kim}
	received the B.S. and M.S. degrees in Civil Engineering from Yonsei University, Seoul, South Korea, in 2005 and 2007, respectively, and the Ph.D. degree in Geodetic Science from The Ohio State University, Columbus, OH, USA, in 2013. He is currently a Staff Scientist with the Roy M. Huffington Department of Earth Sciences, Southern Methodist University, Dallas, TX, USA. His research focuses on investigating Earth surface processes using synthetic aperture radar (SAR) imagery, interferometric SAR (InSAR) techniques, and deep learning. His interests include InSAR applications to understanding wetland hydrology, sinkhole evolution, landslide mechanisms, and the interactions between human activities and surface or subsurface processes.
\end{IEEEbiography}

\begin{IEEEbiography}[{%
	\includegraphics[width=1in,height=1.25in,clip,keepaspectratio]{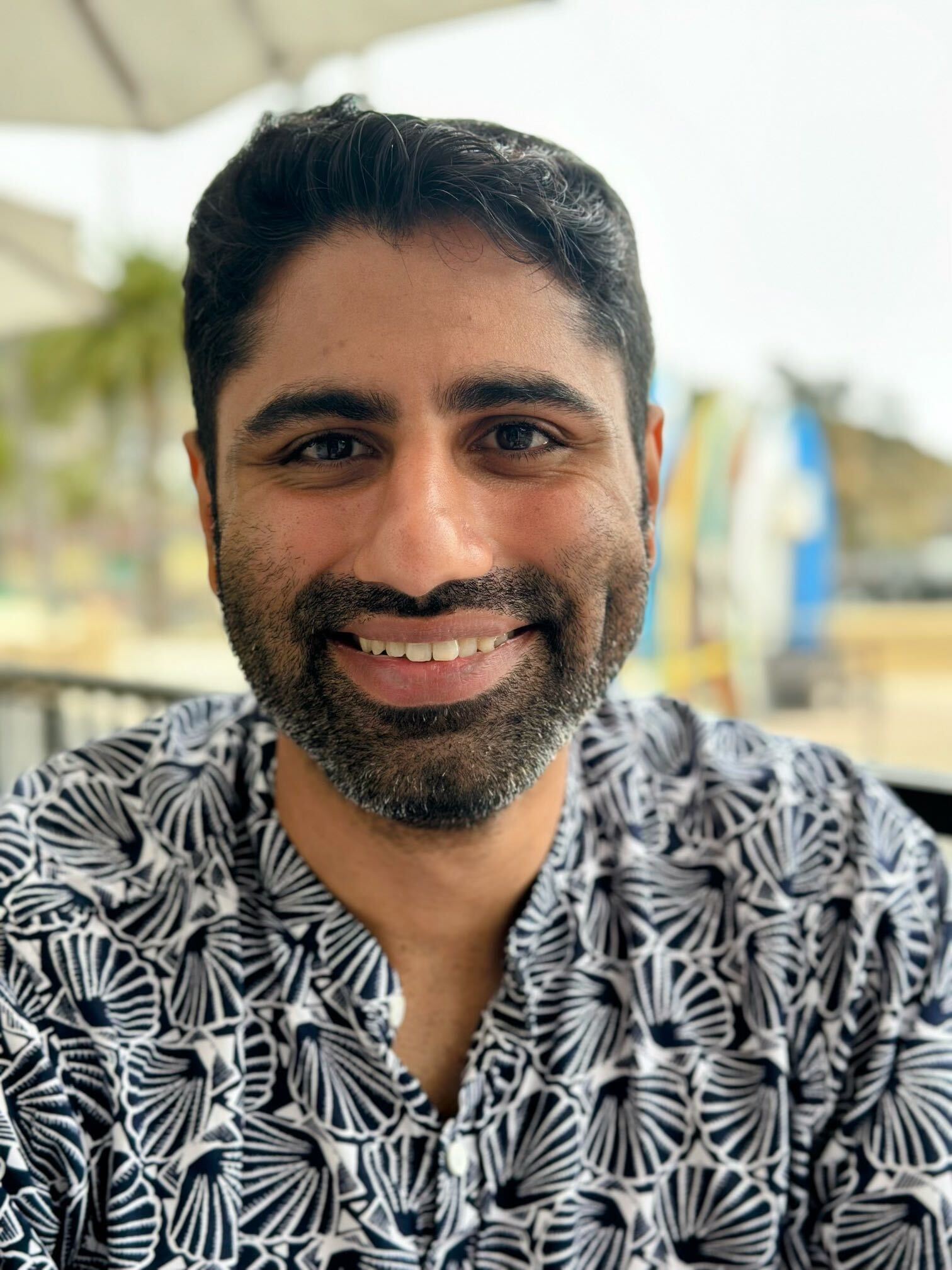}%
	}]{Simran S. Sangha}
	received his B.A. degree in Geology from Occidental College, Los Angeles, CA, USA, in 2014, and his M.S. and Ph.D. degree in Geology in 2017 and 2021, respectively, from the University of California, Los Angeles, CA, USA. He has been working at the Jet Propulsion Laboratory, Pasadena, CA, USA since 2012 when he started interning through the Planetary Geosciences, Earth Surface and Interior, and Radar Science groups, before beginning work full time with the Radar Science group as a Signal Analysis Engineer in 2021. Through his work under the OPERA and ARIA projects, he specializes in using remote sensing techniques to study deformation processes such as those associated with tectonics, hazards, and anthropogenic activities, as well as developing software intended to facilitate the access, management, and use of remote sensing products.
\end{IEEEbiography}

\begin{IEEEbiography}[{%
  \includegraphics[width=1in,height=1.25in,clip,keepaspectratio]{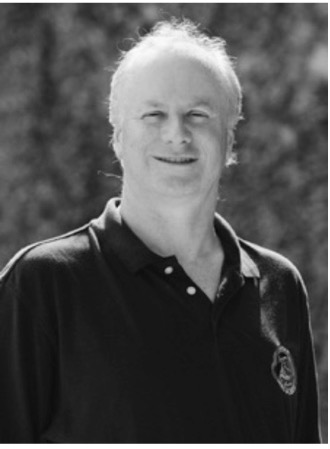}
}]{Bruce D. Chapman}
received his A.B. degree in Physics and in Astronomy from the University of California, Berkeley in 1981 and his Ph.D. from the Department of Earth, Atmospheric, and Planetary Science of the Massachusetts Institute of Technology in Cambridge Massachusetts in 1986.  Since then, he has been a Scientist with the NASA Jet Propulsion Laboratory, California Institute of Technology, Pasadena, California in the Radar Science and Engineering Section.  Currently, he is a member of the NASA-ISRO SAR (NISAR) Science Team, and also leading calibration and validation activities for the NISAR project at JPL.  He is an investigator in the NASA arctic boreal vulnerability experiment (ABOVE), and in JAXA's 3rd research announcement for Earth observation. He is past chair of the SAR subgroup of the CEOS Working Group on Calibration and Validation (WGCV).
\end{IEEEbiography}

\begin{IEEEbiography}[{%
		  \includegraphics[width=1in,height=1.25in,clip,keepaspectratio]{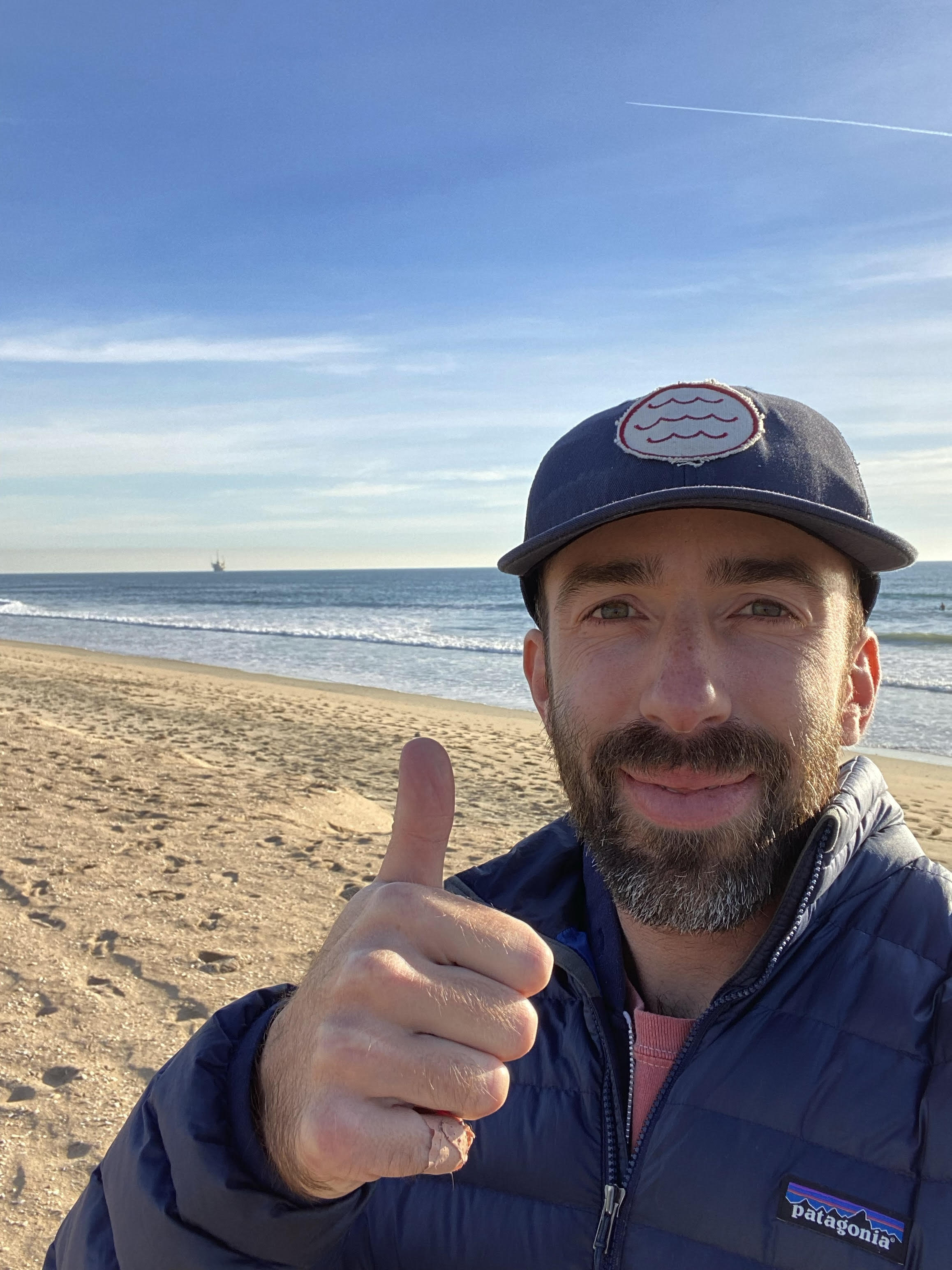}%
	}]{Alexander L. Handwerger}
    received the B.A. degree in Earth Sciences from Boston University, Boston, MA, USA, in 2008, and the Ph.D. degree in Geological Sciences from the University of Oregon, Eugene, OR, USA, in 2015. He is currently a Research Scientist at the Jet Propulsion Laboratory, California Institute of Technology and an Associate Project Scientist at the Joint Institute for Regional Earth System Science and Engineering (JIFRESSE), University of California, Los Angeles, CA, USA. He is the Deputy Project Scientist for the OPERA project. His research interests include landslides, rock glaciers, and satellite and airborne interferometric synthetic aperture radar.
\end{IEEEbiography}

\begin{IEEEbiography}[{%
		 \includegraphics[width=1in,height=1.25in,clip,keepaspectratio]{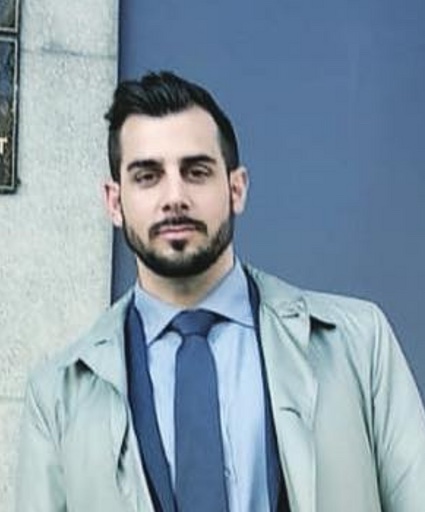}
	}]{Marin Govorcin}
	received the B.Eng. and M.Eng degrees in Geodesy and Geoinformation, in 2013, and the Ph.D. degree in Remote Sensing in 2018., from University of Zagreb, Faculty of Geodesy, Zagreb, Croatia. He is currently a Research Scientist in the Radar Science Group at the Jet Propulsion Laboratory, California, where he leads the Vertical Land Motion (VLM) Product development for the OPERA project. His research interests  include large-scale InSAR time-series processing, uncertainty estimation, geodetic data integration, and solid Earth deformation.
\end{IEEEbiography}

\begin{IEEEbiography}[{%
	\includegraphics[width=1in,height=1.25in,clip,keepaspectratio]{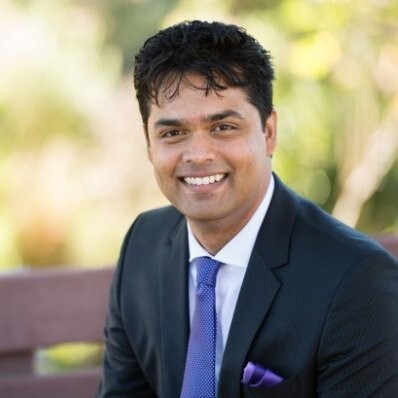}%
	}]{Piyush Agram}
	received the B.Tech. degree in electrical engineering from IIT Madras, Chennai, India, in 2004, and the Ph.D. degree in electrical engineering from Stanford University, Stanford, CA, USA, in 2010. He was a Keck Institute of Space Studies Postdoctoral Scholar with the Caltech's Seismological Laboratory, Pasadena, CA, USA, until 2013 and then joined the Radar Algorithms and Processing Group, Jet Propulsion Laboratory, Pasadena, CA, USA. In 2020, he joined Earthdaily Analytics (formerly known as Descartes Labs) and continues to work there on large scale SAR/InSAR analytics. His research interests include algorithm development for SAR focusing, radar interferometry for deformation time-series applications, and geospatial big data analysis.
\end{IEEEbiography}

\begin{IEEEbiography}[{%
		  \includegraphics[width=1in,height=1.25in,clip,keepaspectratio]{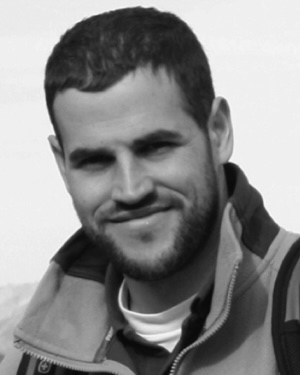}%
	}]{David P. S. Bekaert}
	received his MSc degree in Aerospace Engineering from Delft University of Technology, Delft, The Netherlands, in 2011, and his Ph.D. degree in Geodesy and Geophysics from the University of Leeds, Leeds, U.K., in 2015. In 2011 he worked at the European Space Agency as Young Graduate Trainee on P-band radar clutter suppression algorithms. From 2015 to 2025 he worked as Radar Scientist at the Jet Propulsion Laboratory, California Institute of Technology, Pasadena, CA, USA. While at JPL, he held various leadership roles, including as Principal Investigator for the Advanced Rapid Imaging and Analysis (ARIA) Project and as Project Scientist and Project Manager for the Observational Products for End-Users from Remote Sensing Analysis (OPERA) project overseeing the development of large-scale actionable data products from remote sensing data. In 2025, he joined the Flemish Institute for Technological Research (VITO) where he is currently Copernicus Data Project Manager. Dr. Bekaert is also an independent remote sensing consultant, and a member of the NASA–ISRO Synthetic Aperture Radar (NISAR) Science Team and the NASA Sea Level and Change Team.
\end{IEEEbiography}


\end{document}


\title{Supporting Information for ``Near-Real-Time InSAR Phase Estimation for Large-Scale Surface Displacement Monitoring''}

\author{Scott~Staniewicz,
Heresh~Fattahi,
Sara~Mirzaee,
Talib~Oliver-Cabrera,
Emre~Havazli,
Geoffrey~Gunter,
Se-yeon~Jeon,
Mary~Grace~Bato,
Simran~Sangha,
Alexander~Handwerger,
Marin~Govorcin,
Piyush~Agram,
Jin-Woo~Kim,
David~Bekaert
}


\maketitle

\setcounter{section}{0}
\setcounter{figure}{0}
\setcounter{table}{0}
\setcounter{equation}{0}

\renewcommand{\thesection}{S\arabic{section}}
\renewcommand{\thefigure}{S\arabic{figure}}
\renewcommand{\thetable}{S\arabic{table}}
\renewcommand{\theequation}{S\arabic{equation}}


\section{Regridding PS and DS to 30 m resolution}
\label{sec:supplement-regridding}

\begin{figure}[!hbt]
	\centering
	\includegraphics[width=0.95\columnwidth]{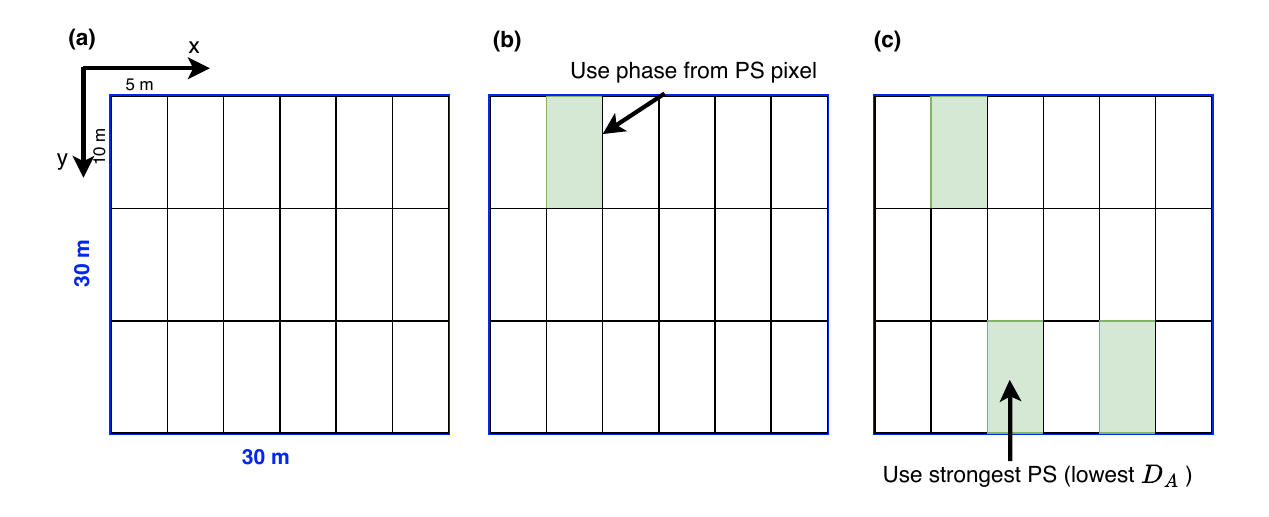}
	\caption{(a) Illustration of one output DISP-S1 pixel compared to input CSLC-S1 grid
		(b) Output phase selection for the single-PS case
		(c) Output pixels with multiple PS use the strongest (lowest $D_A$) phase}
	\label{fig:ps-regridding}
\end{figure}

After estimating the wrapped phases for both DS and PS, interferograms are created at a coarser resolution.  Our input SLC products are geocoded on a Universal Transverse Mercator (UTM) grid with 5 meter $\times$ 10 meter pixel posting in X- and Y-directions, respectively, while the interferograms are created at 30 m x 30 m pixel posting (Figure \ref{fig:ps-regridding} (a)). Note that while the PS processing occurs at full SLC resolution, the DS processing, where a multi-looked covariance matrix is processed using phase linking, occurs only once per output grid pixel to speed up the processing by a factor of 18. Since there are 18 full-resolution SLC pixels are contained within each output interferogram pixel, there are three possible cases for setting the interferogram phase:
\begin{enumerate}
	\item \emph{The decimation window contains no PS pixels}: In this case, the DS phase linking solution from the center of the decimation window is used.
	\item \emph{One PS is present in the decimation window}: the single-reference PS phase is used (Figure \ref{fig:ps-regridding} (b)).
	\item \emph{Multiple PS pixels are present}: Choose the phase from the PS pixel with the lowest $D_A$ (Figure \ref{fig:ps-regridding} (c)).
\end{enumerate}

\section{Compressed SLC strategy for separate batched processing}
\label{sec:supplement-sequential-last-per-ministack}

The sequential processing algorithm outlined in Section IIB, Main Text, produces high quality wrapped phase estimates at a fraction of the computational cost as forming full covariance matrices. When intermediate outputs are available to future mini-stacks, any desired interferograms may be re-formed after phase linking. As an example using Figure 1c, we can form the interferogram with reference date 8 to secondary day 9, by cross multiplying the phase linking output from day 1 to day 8 with the output from day 1 to day 9.  However, if the data processing system does not have access to the phase linking outputs from previous mini-stacks, we cannot form the interferogram from day 8 to day 9; we only have access to compressed SLCs with reference phase from day 1. 

\begin{figure*}[!ht]
	\centering
	\includegraphics[width=0.95\textwidth]{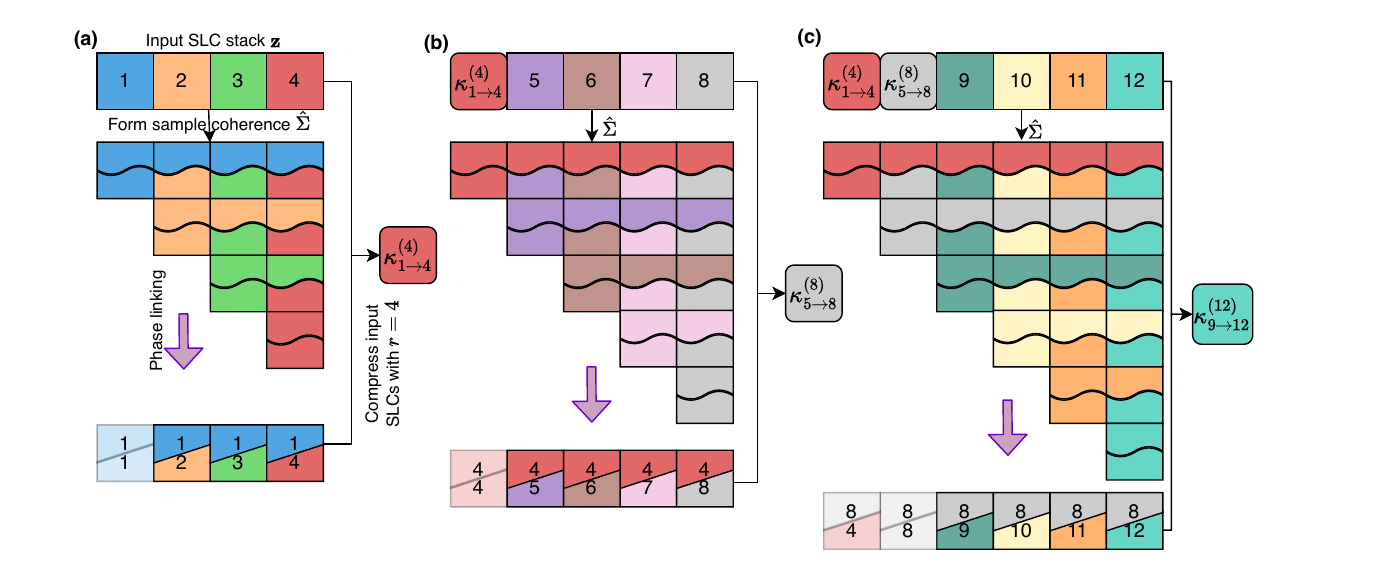}
	\caption{Sequential phase linking scheme used in OPERA processing to create short-temporal baseline mini-stacks with separate processing runs.
		(a) The first mini-stack (images 1-4) is phase-linked, and input SLCs are compressed with reference index $r=4$ to produce $\compslc{4}{1}{4}$.
		(b) The second mini-stack uses $\compslc{4}{1}{4}$ as a reference, which allows the formation of an interferogram between day 4 and day 5 (as opposed to day 1 and day 5 in Figure 1 of the Main Text).
		(c) Each subsequent mini-stacks uses phase from the latest real SLC to create new compressed SLCs.
	}
	\label{fig:sequential-demo-last-per-ministack}
\end{figure*}

Since unwrapping long temporal baseline interferograms may lead to poor results, we can adapt the compressed SLC scheme to move the reference date forward for each new compressed SLC (Figure \ref{fig:sequential-demo-last-per-ministack}). Instead of using the most recent Compressed SLC as the reference phase (c.f. Figure 1, main text) we use the phase from the \emph{last real SLC} of each mini-stack. The benefit is that in subsequent mini-stacks, we have the ability to unwrap interferograms with shorter temporal baselines; the downside is that the final displacement outputs are unable to be referenced to the first date of the archive without using previously archived displacement results. For the OPERA DISP-S1 product, this tradeoff was deemed worthwhile to get the highest quality results for rapid displacement events (such as volcanic eruptions and landslides), as users who wish to reconstruct full displacement time series can be daisy-chain consecutive mini-stack results together using the overlapping reference dates.

\section{Sentinel-1 acquisition selection and exclusion strategy}
\label{sec:supplement-exclude-bursts}
Since the input CSLC-S1 products were produced separately for each Burst ID, the output DISP-S1 frames do not need to follow the same frame boundaries as the ESA-produced Level-1 SLC products. This allows the DISP-S1 product to maintain a consistent spatial coverage over time so users to not need to filter out partial frames.

\begin{figure*}[!ht]
	\centering
	
    \includegraphics[width=0.55\textwidth]{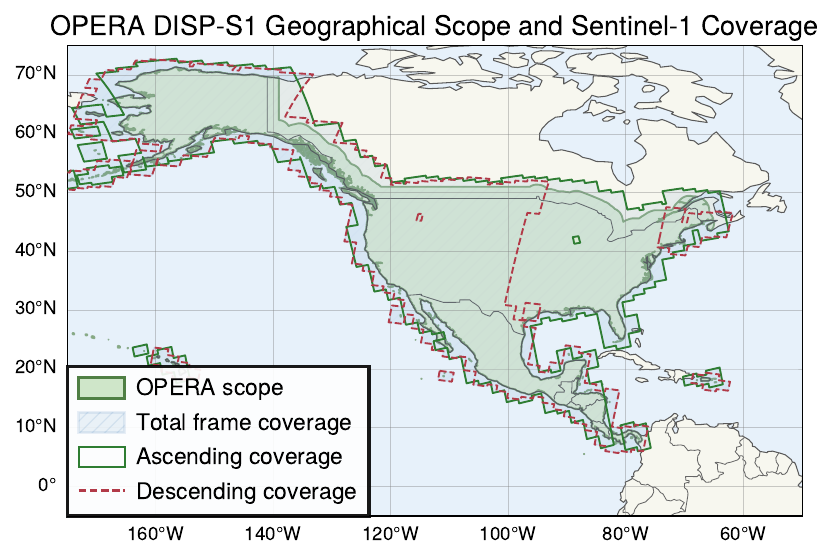}
	\caption{OPERA geographical scope for North America, defined as the U.S. and U.S. Territories, Canada within 200 km of the U.S. border, and all mainland countries from the southern U.S. border down to and including Panama (green). DISP-S1 Frame coverage (light blue), defined from ESA's Burst ID map, covers this region, even for frames which contain no acquisitions. The ascending frames (green outline) contain more complete data than the descending frames (red dotted). }
	\label{fig:supplement-opera-scope}
\end{figure*}

To ensure consistent spatial coverage for all dates included all fixed frames, we analyzed the historical coverage of Sentinel-1 available between July, 2016 and December, 2024 (Figure \ref{fig:supplement-opera-scope}). The goal of the analysis was to select a subset of the 27 Burst IDs contained within each nominal DISP-S1 frame such that (1) the data included on each date has the same spatial coverage, and (2) the maximum number of total CSLC-S1 products are used. For example, if a frame is missing Sentinel-1 acquisitions for 15 of the Burst IDs on a single date, that date is ignored; however, if a frame is missing 15 of the burst IDs for nearly all dates, those burst IDs are excluded (\textit{i.e.}, a spatial gap is created in the data).

\section{Seasonal snow analysis and exclusion strategy}
\label{sec:supplement-snow}
To reduce the impact of decorrelation on regions with seasonal snow cover, we implemented a ``snow blackout'' strategy to systematically exclude winter acquisitions. Seasonal snow and persistent sub-freezing surface temperatures can lead to reduced temporal coherence in InSAR time series, with snowpack and moisture changes rather than geophysical deformation.

We used the National Oceanic and Atmospheric Administration's (NOAA) Global Ensemble Forecast System (GEFS) reanalysis data spanning June, 2020 - June, 2025. We focused on the 2 meter air temperature and categorical surface snow cover variables to find adverse environmental conditions. Data were aggregated to find the maximum daily mean temperature, and the total number of days in a week marked with snow. Each model grid cell was flagged as ``bad'' for that week if either (1) the weekly maximum average temperature was below $-2^\circ$C, or (2) the pixel had 4 or more days of snow cover during the week. For each DISP-S1 frame, a week was considered unusable if more than 50\% of the frame's pixels were flagged as ``bad.'' Finally, the 5 years of data were aggregated to produce a single blackout start and end date to be used each year in production for simplicity.

For each frame, weekly ``exclusion'' flags were generated for each grid cell of the GEFS model for the 5-year period from July, 2020 to July, 2025. The flags were joined to the DISP-S1 frame locations to estimate a single blackout start and end date to be used each year in production. Since the five years contain variability, three candidate blackout strategies were defined:

\begin{enumerate}
	
	\item Shortest blackout period: latest observed onset of winter and earliest observed spring melt were used as the start and end dates.
	
	\item Longest blackout period: earliest observed onset of winter and latest observed melt.
	
	\item Median onset and melt dates across the five years.
\end{enumerate}
The median strategy was selected as the default for the majority of frames. If the median blackout exceeded a threshold duration or 210 days ($\sim$ 7 months) for a given frame, the shorter blackout period was used for that frame to retain more data. Frames containing high-priority volcanoes identified in the 2018 USGS National Volcanic Threat Assessment \cite{ewert20182018} were assigned the conservative strategy to ensure that snow-covered volcanic summits were excluded even if surrounding lowlands were snow-free. For example, over the Three Sisters volcano (Frame 30710), the blackout period begins on November 3rd and ends on May 15th, resulting in exclusion of all acquisitions within that range.

In addition to excluding winter periods with snow, the rainiest 4-month period (August, September, October, and November) for Central America was excluded due to strong decorrelation observed from heavy vegetation and moisture changes.

\begin{figure*}[!ht]
	\centering
    \includegraphics[width=0.95\textwidth]{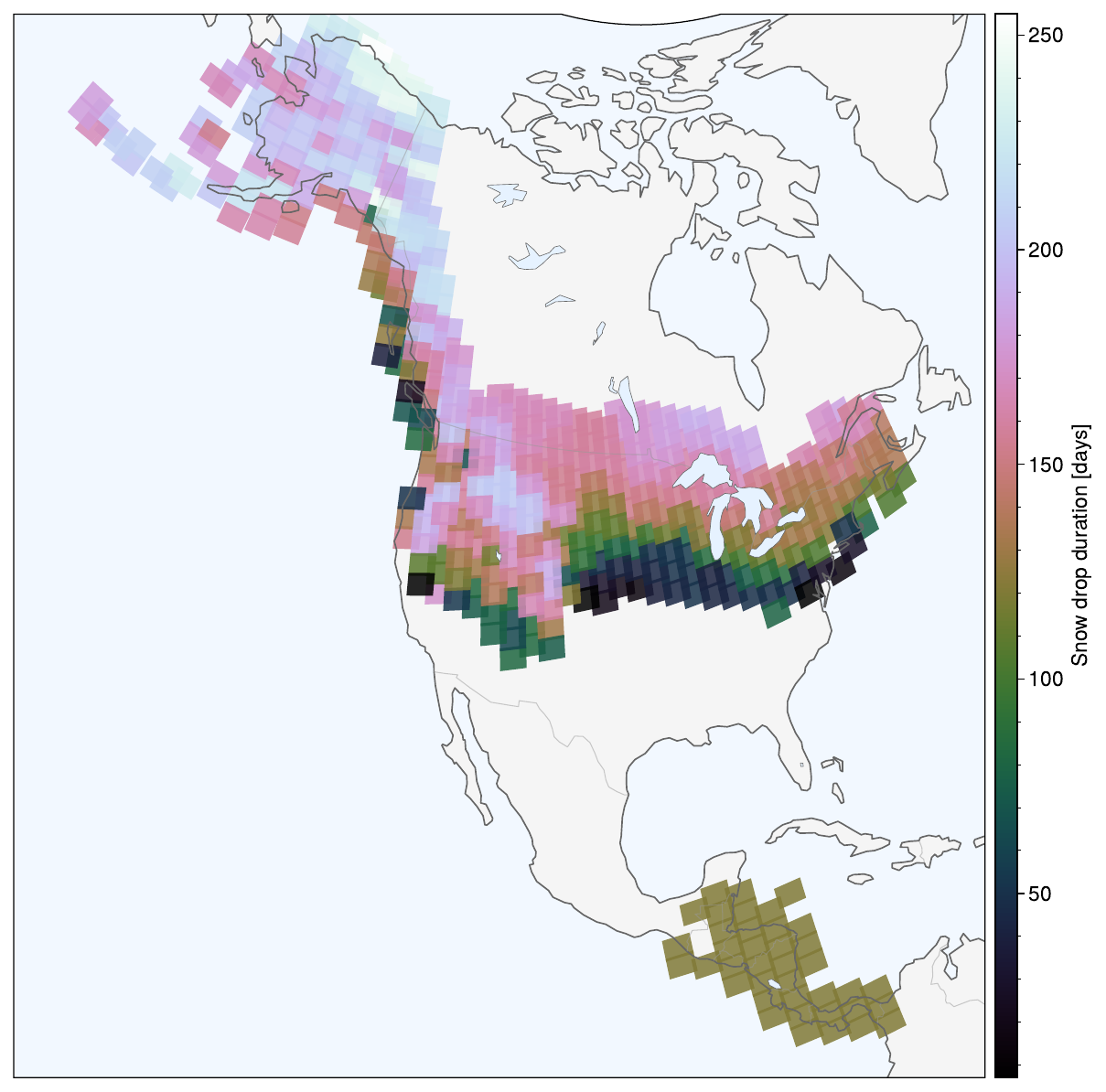}
	\caption{Durations (in days) of excluded time periods for ascending frames. Winter months are excluded northern latitudes, and rainy months (Aug. -- Nov.) are excluded for Central America.}
	\label{fig:supplement-snow-drop-strategy}
\end{figure*}

A summary of blackout durations across ascending and descending passes is shown in Figure \ref{fig:supplement-snow-drop-strategy}.

\newpage
\section{SHP window size configuration by region}
\label{sec:supplement-shp-window}

The SHP search window size was configured per geographic region based on empirical evaluation of processing quality, including unwrapping error rates and temporal coherence. Regions were defined by geographic polygons and translated to OPERA Sentinel-1 Frame IDs. Table~\ref{tab:shp-window-sizes} lists the half-window and full window sizes used for each region. The GLRT adaptively selects only statistically homogeneous pixels within the search window, so the effective averaging neighborhood is typically smaller than the full window; the per-pixel count of selected SHP neighbors is recorded in the \texttt{shp\_counts} layer of each DISP-S1 product. The regional override configurations, including the geographic polygons defining each region, are publicly available in the processing software repository.

\begin{table}[!htbp]
	\centering
	\caption{SHP search window sizes by geographic region. Half-window values define the search radius in the Y (azimuth) and X (range) directions; the full window is $(2h+1)$ pixels in each dimension. Physical sizes assume the 5\,m $\times$ 10\,m CSLC-S1 pixel posting.}
	\label{tab:shp-window-sizes}
	\begin{tabular}{l c c c}
		\toprule
		Region & Half-window (x $\times$ y) & Full window (X $\times$ Y) & Approx.\ size (m) \\
		\midrule
		Low-vegetation west & $12 \times 6$ & $25 \times 13$ & $\sim$125 $\times$ 130 \\
		N.\ California--Cascadia & $18 \times 9$ & $37 \times 19$ & $\sim$185 $\times$ 190 \\
		Alaska (interior) & $18 \times 9$ & $37 \times 19$ & $\sim$185 $\times$ 190 \\
		Alaska (North Slope) & $19 \times 9$ & $39 \times 19$ & $\sim$195 $\times$ 190 \\
		Remaining & $16 \times 8$ & $33 \times 17$ & $\sim$165 $\times$ 170 \\
		\bottomrule
	\end{tabular}
\end{table}

\newpage
\section{Description of Validation Sites}
\label{sec:supplement-validation_sites}
Table \ref{tab:disp_s1_validation_sites} summarizes the validation sites and thematic areas selected for assessing the accuracy of the OPERA DISP-S1 product. The sites span diverse geophysical and environmental settings across North America, encompassing both negligibly deforming and highly deforming regions influenced by natural and anthropogenic processes. Each site is associated with a defined validation approach (VA1 or VA2) and characterized by its climate and surface conditions. This selection ensures that the validation framework represents the range of environmental variability, deformation behaviors, and climate conditions expected within the North America geographic scope of the project.

\begin{table*}[!htbp]
    \centering
    \caption{Overview of validation sites, applied approaches, thematic areas and associated climatic and environmental characteristics used in evaluating the accuracy of the OPERA DISP-S1 product.}
    \label{tab:disp_s1_validation_sites}
    \begin{tabular}{|>{\centering\arraybackslash}p{2.3cm}|
                >{\centering\arraybackslash}p{2.3cm}|
                >{\centering\arraybackslash}p{2.5cm}|
                >{\centering\arraybackslash}p{2.8cm}|
                >{\centering\arraybackslash}p{2.8cm}|}
    \hline
    \textbf{Thematic Area} & 
    \textbf{Validation Sites} & 
    \textbf{Validation Approach} & 
    \textbf{Climate} & 
    \textbf{Characteristics} \\
    \hline
    Tectonic deformation, Local Land Subsidence (natural and anthropogenic), Landslides	& California (F11115, F11116, F36542, F18903, F09156) & VA1, VA2. Frame F11116 was excluded from VA2 as most of the area exhibits significant deformation. & Temperate/Dry/ Hot & Combination of rural and urban landscapes with diverse topography and decorrelation sources; varying base elevations across frames. \\
    
    \hline
    Volcano deformation	& Big Island, Hawaii (F33039) & VA1, VA2 & Tropical/Rainforest, Tropical/Savanna & Predominantly urban and rainforest environment on a tropical volcanic island; variable relief and strong decorrelation from vegetation and lava flows; active volcano inflation and deflation processes. \\
    
    \hline
    Stable/mostly non-deforming deformation	& Oklahoma (F28486) & VA1, VA2 & Temperate/Dry/ Hot & Agricultural region, strong atmosphere, flat/no relief. \\
    
    \hline
    Local land subsidence (anthropogenic and natural sources) & Houston-Galveston, Texas (F08882) & VA1, VA2 & Subtropical/Wet/Hot & Urbanized coastal plain environment characterized by flat relief, wetlands, hydrocarbon fields, and subsurface salt domes; subject to anthropogenic and natural subsidence including sinkhole development; humid subtropical Southern US climate. \\
    
    \hline
    Volcano deformation, Tectonic deformation & Unimak Island, Alaska (F33065) & Only VA2; insufficient high-quality GNSS coverage in the area. & Cold/Wet/Subarctic & Volcanic ocean island exhibiting strong topographic relief and unstable atmospheric condition; deformation and coherence may be affected by seasonal snow accumulation and freeze-thaw transitions. \\
    
    \hline
    Landslides, Glacier dynamics & South Central Alaska (F42779) & Only VA2; insufficient high-quality GNSS coverage in the area. & Cold/Wet/Subarctic & Complex topography spanning steep coastal mountains, fjords, alpine zones, and lowland basins; Actively adjusting paraglacial landscape. \\
    
    \hline
    Local Land Subsidence & New York (F08622) & VA1, VA2 & Hot/Cold/Subtropical &  Varied terrain ranging from coastal lowlands to upland mountains, dominated by deciduous and mixed conifer forests; surface conditions can be impacted by seasonal snowpack, freeze–thaw cycles, and vegetation seasonality. \\
    
    \hline
    Local Land Subsidence & Central Florida (F12640) & Only VA2; insufficient high-quality GNSS coverage in the area. & Hot/Wet &  Low-elevation plains with abundant lakes and wetlands, karstic subsidence features, and strong seasonal moisture variability. \\
    
    \hline
    \end{tabular}
\end{table*}

\newpage
\section{Descending velocity mosaic and quality metric mosaics}
\label{sec:supplement-mosaics}

\begin{figure*}[!ht]
	\centering
	\includegraphics[width=0.99\textwidth]{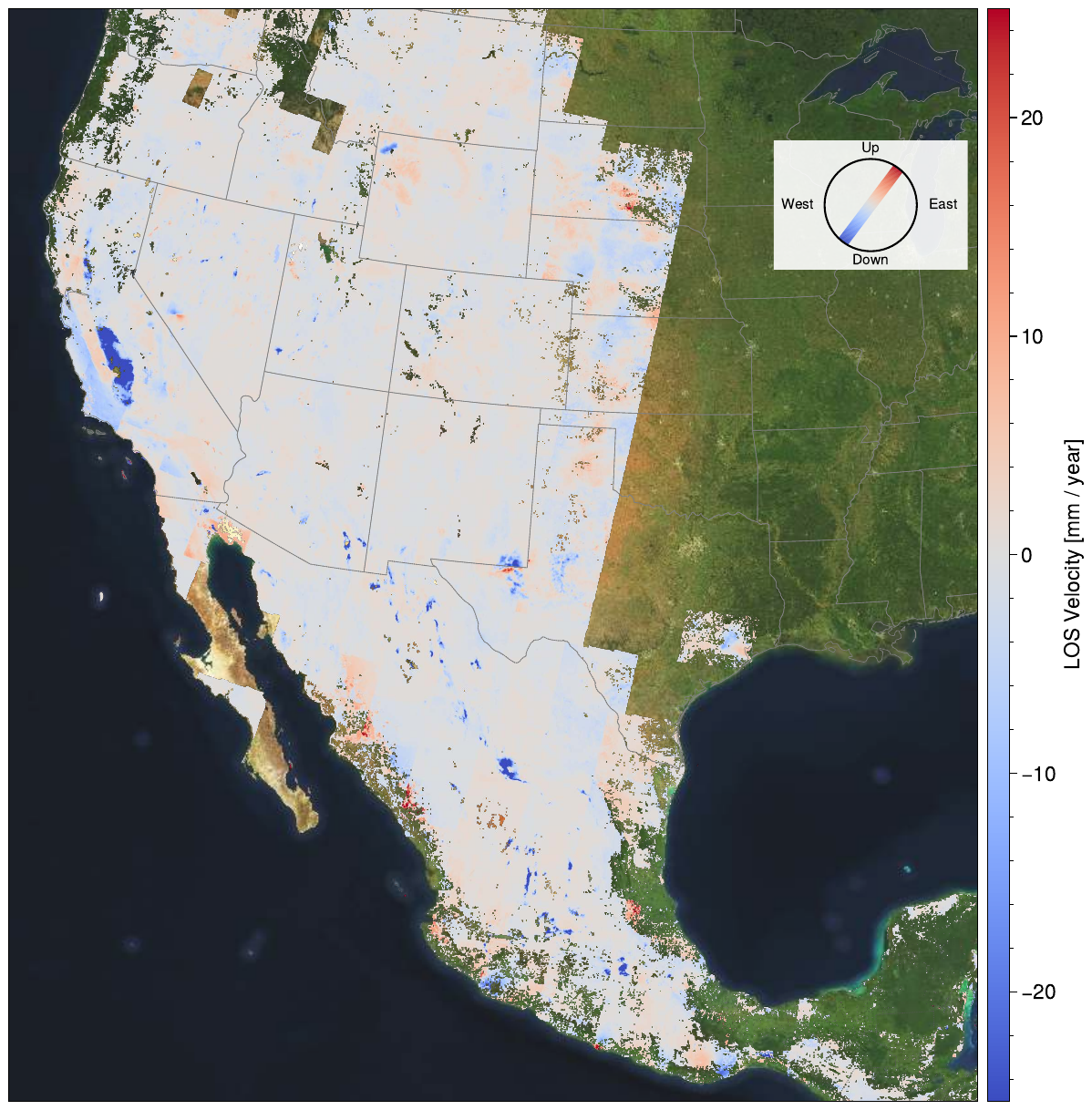}
	\caption{Average LOS surface velocity from Nov. 2017 to Nov. 2024 for the descending Sentinel-1 geometry. Red indicates average motion toward the satellite, blue indications motion away. Velocities are relative to an internal average rate per frame. For wide-area visualization continuity, a planar ramp was removed from each frame, and a constant shift was estimated for each frame using along-track overlapping regions.}
	\label{fig:supplement-descending-conus-7year-velocity}
\end{figure*}

\begin{figure*}[!ht]
	\centering
	\includegraphics[width=0.99\textwidth]{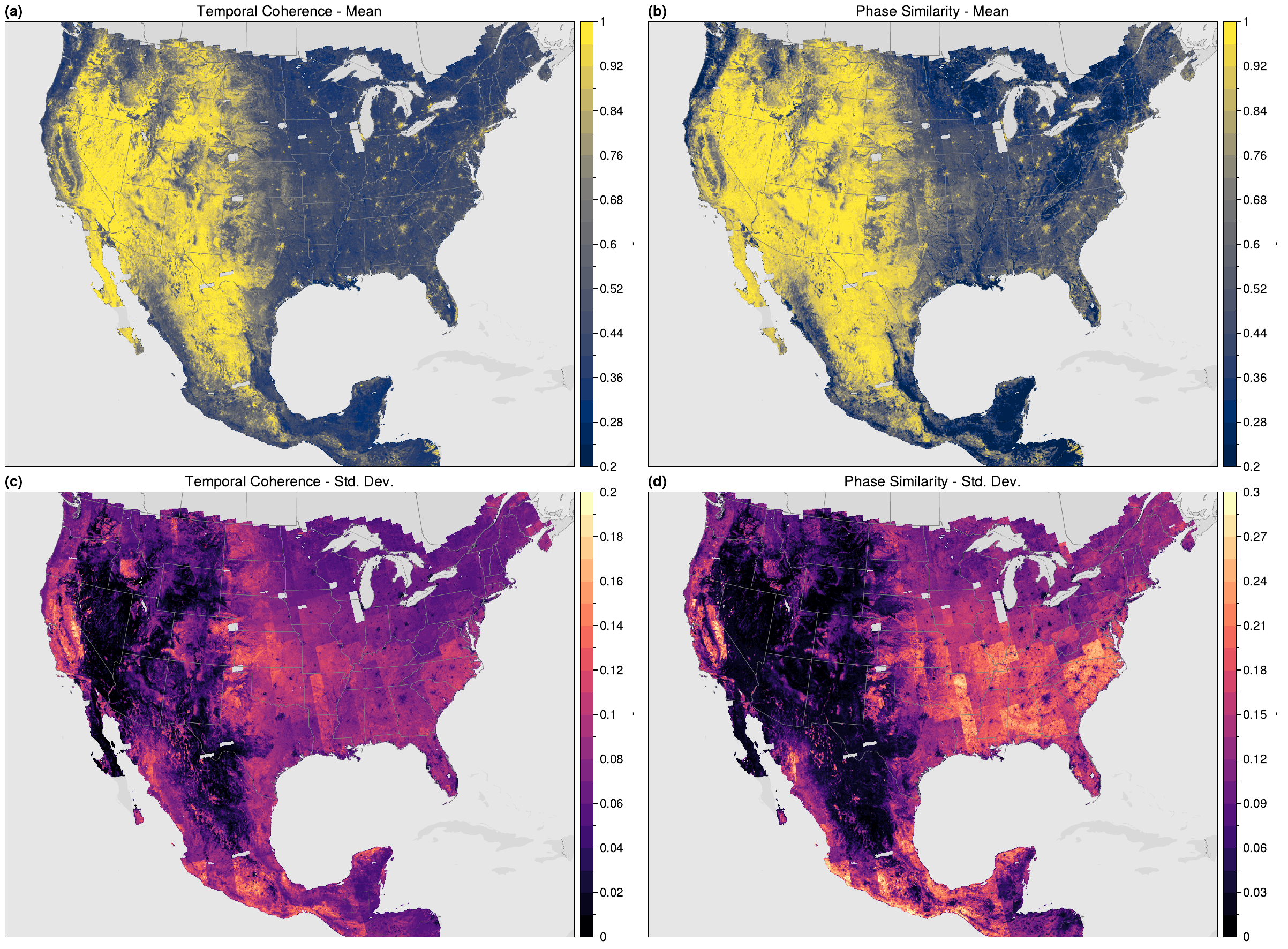}
	\caption{Ascending quality metrics for (right column) temporal coherence and (left) phase similarity, showing mean (top row) and standard deviation (bottom row) for the 7 year period from Nov. 2017 to Nov. 2024. }
	\label{fig:supplement-ascending-conus-quality}
\end{figure*}

\IEEEpeerreviewmaketitle

\bibliographystyle{IEEEtran}
\bibliography{IEEEabrv,citations}

\ifCLASSOPTIONcaptionsoff
  \newpage
\fi